\documentstyle[emulateapj]{article}

\def\etal{{\rm et~al.\ }}
\def\sdm{\sigma_{\rm DM}}
\def\hmpc{\;h^{-1}{\rm Mpc}}
\def\Mpc3{\;M_\odot {\rm pc}^{-3}}

\def\hkpc{h^{-1}{\rm kpc}}
\def\kms{{\rm \;km\;s^{-1}}}

\def\kmsmpc{\kms\;{\rm Mpc}^{-1}}
\def\xunits{{\rm \;cm^2}{\rm GeV}^{-1}}

\def\r200{r_{\rm 200}}
\def\v200{v_{\rm 200}}
\def\cnfw{c_{\rm NFW}}
\newcommand{\PSbox}[3]{\mbox{\rule{0in}{#3}\includegraphics{#1}\hspace{#2}}}

\lefthead{Dav\'e, Spergel, Steinhardt, Wandelt} 
\righthead{Cosmological Simulations of SIDM}

\begin{document}

\title{Halo Properties in Cosmological Simulations of Self-Interacting Cold Dark Matter}

\author{
Romeel Dav\'{e}$^{1,3}$, David N. Spergel$^{1}$, 
Paul J. Steinhardt$^{2}$, Benjamin D. Wandelt$^{2}$
} 
 
\footnotetext[1]{Princeton University Observatory, Princeton, NJ 08544}
\footnotetext[2]{Department of Physics, Princeton University, Princeton, NJ 08544}
\footnotetext[3]{Spitzer Fellow}

\begin{abstract} 

We present a comparison of halo properties in cosmological simulations
of collisionless cold dark matter (CDM) and self-interacting dark
matter (SIDM) for a range of dark matter cross sections.  We find, in
agreement with various authors, that CDM yields cuspy halos that are
too centrally concentrated as compared to observations.  Conversely,
SIDM simulations using a Monte Carlo $N$-body technique produce halos
with significantly reduced central densities and flatter cores with
increasing cross section.  We introduce a concentration parameter based
on enclosed mass that we expect will be straightforward to determine
observationally, unlike that of Navarro, Frenk \& White, and provide
predictions for SIDM and CDM.  SIDM also produces more spherical halos
than CDM, providing possibly the strongest observational test of SIDM.
We discuss our findings in relation to various relevant observations as
well as SIDM simulations of other groups.  Taking proper account of
simulation limitations, we find that a dark matter cross section per
unit mass of $\sdm\approx 10^{-23}-10^{-24}\xunits$ is consistent with
all current observational constraints.
\end{abstract}
 
 
\section{Introduction}

The cold dark matter (CDM) family of cosmological models provides an
excellent description of a wide variety of observational results, from
the earlier observable epochs detected via microwave background
fluctuations to present-day observations of galaxies and large-scale
structure.  A ``concordance model" with roughly one-third matter and
two-thirds vacuum energy, either a cosmological constant or 
quintessence (\cite{cal98}), is consistent with almost all current
observations on scales $\ga 1$~Mpc (\cite{bah99}).

Recently, improving observations and numerical techniques have enabled
a comparison of CDM scenarios to observations on galactic scales
of $\sim$~few~kpc.  The results have not been encouraging.  There are a
number of distinct observations that may be in conflict with
predictions of CDM:
\begin{itemize}

\item The density profile of galaxies in the inner few kiloparsecs
appears to be much shallower than predicted by numerical simulations
(Navarro, Frenk \& White 1996, hereafter \cite{nfw96}).  For density
profiles characterized by $\rho(r)\propto r^\alpha$ as $r\rightarrow
0$, CDM predicts $\alpha\approx -1.5$ with little scatter (\cite{moo99}
1999), while current H$\alpha$ observations suggest $\alpha\approx
-0.5$ with significant scatter (\cite{swa00}; \cite{dal00}, though
see \cite{vs00}).
\item The central density of dark matter halos is observed to be
$\rho_c\sim 0.02 \Mpc3$ roughly independent of halo mass
(\cite{fir00b}), while CDM predicts halos with $\rho_c\ga 1 \Mpc3$ at
dwarf galaxy masses, increasing to larger masses (\cite{moo99} 1999).
\item The number of dwarf galaxies in the Local Group is an order of magnitude 
fewer than predicted by CDM simulations, with the discrepancy growing 
towards smaller masses (\cite{moo99a}; \cite{kly99}).
\item Hydrodynamic simulations produce galaxy disks that are too small and
have too little angular momentum, yielding a Tully-Fisher relation whose
zero-point is off by several magnitudes from observations (\cite{nav00}).
\item The robustness of rapidly rotating bars in high surface brightness 
spiral galaxies implies lower density cores than predicted by CDM (\cite{deb98}).
\item Cluster CL~0024+1654 is nearly spherical with a large, soft core,
while CDM typically predicts triaxial clusters with cuspy cores (\cite{tys98},
though see Miralda-Escud\'e 2000 for a counterexample).
\end{itemize}

Each piece of evidence taken individually is perhaps not convincing
enough to claim that CDM has failed on galactic scales.  For instance,
until recently there was controversy amongst simulators regarding
inner profiles (\cite{kra98}), but more careful simulations have
converged on a consistent prediction (\cite{kly00}).  Observationally,
inner galactic profiles are uncertain due to beam smearing effects in
\ion{H}{1} observations (\cite{swa00}; \cite{vdb00}), though samples
of high-resolution H$\alpha$ observations continue to show shallower
profiles than predicted by CDM (\cite{dal00}).  The number of observed Local
Group dwarf galaxies may be reconciled with CDM via plausible scenarios
for suppressed galaxy formation (e.g.  \cite{bul00}), or else compact
high-velocity clouds could represent the ``missing satellites" that are
seen in $N$-body simulations (\cite{bli99}).  Hydrodynamic simulations
of disk galaxy formation are fraught with the usual concerns about the
effects of feedback, artificial viscosity and resolution, though it
appears the discrepancies above are due to the underlying dark matter
distribution (\cite{nav00}).

It becomes more interesting to consider alternatives to conventional
CDM when one recognizes that all these discrepancies may be symptomatic
of a single cause:  {\it Dark matter halos in CDM simulations appear to
be more centrally concentrated than observed.} Recognizing this,
various authors have recently forwarded a plethora of alternative dark
matter theories that suppress the central concentration of dark matter
in galaxy halos.  Among such theories are that the dark matter is warm
(\cite{som00}; \cite{col00}; \cite{han00}), repulsive (\cite{goo00}),
fluid (\cite{pee00}), fuzzy (\cite{hu00}), decaying (\cite{cen00}),
annihilating (\cite{kap00}), and the alternative we investigate here,
self-interacting (SIDM; Spergel \& Steinhardt 2000).  Interestingly,
all theories may be tuned to solve the problems mentioned above (at
least in analytic approximations), all theories may be motivated from
particle physics considerations, and all theories retain the desirable
properties of CDM on extragalactic scales (though warm dark matter is
non-trivially constrained by this requirement; see \cite{nar00}).

SIDM is governed by a single free parameter, the cross section per unit
mass $\sdm$ of the interacting dark matter particle.  \cite{spe00}
suggested $\sdm\sim 10^{-22}-10^{-25}\xunits$ in order to reduce the
central concentration of galaxy halos by a sufficient amount to
alleviate the above problems.  Intriguingly, this value is close to the
cross section of ordinary hadrons, motivating some particular particle
physics candidates for SIDM (Steinhardt et al., in preparation).  If
$\sdm$ is significantly smaller than this range, then the optical depth
at galactic densities is much less than unity, implying that SIDM would
have a negligible effect on the dark matter distribution in halos.  

A qualitative picture of the evolution of an SIDM halo is as follows:
At early times there is no difference between SIDM and CDM since the
densities and peculiar velocities are sufficiently low  that collisions
are rare; hence SIDM makes identical predictions to CDM regarding
cosmic microwave background fluctuations and the Lyman alpha forest.
As the halo forms and grows via gravitational instability, the central
density increases.  Eventually, collisions are so frequent that dark
matter particles scatter out of the center as fast as they are
accreted, and the density growth is halted, forming a core.  Such a
limit is not present in the CDM model, where the central density grows
unchecked.  The SIDM core then begins to extend while retaining
constant central density. Heat transfer from the outer parts of the
halo raises the temperature in the halo core. If the halo is truly
isolated, then eventually the core thermalizes with the exterior
resulting in an isothermal halo with a steep density profile. This
initiates gravothermal collapse, where the direction of heat transfer
is reversed and the exterior begins to cool the halo center.  However,
in a realistic cosmological setting, galaxies constantly accrete
material, keeping the outer halo hot and heat flowing inwards, thus
delaying core collapse.  The interplay between collisional heat
transfer and accretion determines whether a halo will undergo core
collapse in a Hubble time.

It is important to appreciate that the transport behavior does not
change  monotonically with $\sdm$.  For small cross-sections, heat
transfer increases with $\sdm$ since the frequency of collisions
increases;  however, for large cross-sections, the conductivity
$\kappa\propto\sdm^{-1}\rightarrow 0$ and no heat transfer occurs.
Thus, as we discuss in \S\ref{sec: comparison}, the fluid approximation
is a poor decription of SIDM in the moderate cross-section regime
proposed by Spergel and Steinhardt.  Furthermore,  the behavior in the
moderate cross-section regime cannot be surmised by interpolating
between the fluid  and the non-interacting CDM regimes.  A proper
treatment of the SIDM proposal, which includes  the interplay of
accretion and heat transfer, its non-monotonic dependence $\sdm$, and
the effects of merging demands numerical simulations designed to
explore the moderate cross-section regime.

In this paper we investigate the statistical properties of halos in SIDM
and CDM in cosmological $N$-body simulations.  Our spatial and mass
resolutions are sufficient to probe the inner regions ($\sim 1\hkpc$)
of small halos ($\ga$~few$\times 10^{9} M_\odot$), while maintaining sufficient
volume so as to have a significant sample of such halos.  We use a Monte
Carlo technique similar to \cite{koc00} (2000) to model collisions.
The primary difference between our simulations and prior investigations
(discussed in more detail in \S\ref{sec: comparison}) is that we
model a cosmologically-significant random volume of the universe with
self-interaction cross sections in the range favored by \cite{spe00},
enabling us to characterize the statistical properties of halos as we
vary $\sdm$.

\S\ref{sec: sim} describes our initial conditions and simulation
techniques using a Monte Carlo $N$-body approach.  In \S\ref{sec:
structure} we compare the structural properties of halos in CDM versus
SIDM models with several cross sections.  In particular we examine
their central densities, inner profile slopes $\alpha$, the mass
dependence of $\alpha$, concentrations, phase space densities, and
ellipticities, and where possible compare to observations.  In
\S\ref{sec:  res} we use lower resolution simulations to test the
effects of finite particle numbers in our Monte Carlo method.  In
\S\ref{sec:  subhalo} we examine the subhalo population around the
largest halo in our simulations.  In \S\ref{sec:  comparison} we
compare our findings to the simulations of other groups who have
conducted numerical studies of self-interacting dark matter, and
examine results from a wider range of $\sdm$.  We summarize our results
and discuss observational constraints in \S\ref{sec: disc}.  We find
that SIDM with $\sdm\approx 10^{-23}-10^{-24}\xunits$ produces halos
that are in better agreement than collisionless CDM for a wide variety
of observations.

\section{Simulating SIDM}\label{sec: sim}

\subsection{Code and Cosmology}

We use a modified version of GADGET (\cite{spr00}), a publically-available
TreeSPH code for distributed-memory parallel machines.  Here we only
employ the gravitational $N$-body portion.  We evolve a $4\hmpc$
randomly-chosen volume of a $\Lambda$CDM universe with $\Omega=0.3$,
$\Omega_\Lambda=0.7$, $H_0=70\kmsmpc$, and $\sigma_8=0.8$, similar
to the ``concordance model" in agreement with a wide variety of
observations (\cite{bah99}).  We generate initial conditions using
COSMICS (\cite{ma95}) at $z=49.7$, where our particle distribution
first becomes nonlinear, and evolve to $z=0$.  We employ $128^3$
dark matter particles in each run, resulting in a dark matter particle
mass of $m_p=3.6\times 10^6 M_\odot$, and a spline kernel softening of
$\epsilon=1\hkpc$ (i.e. force is Newtonian at $2\epsilon$).  To test
resolution effects, we also run a suite of simulations with $64^3$
particles and $\epsilon=2\hkpc$.  Their initial conditions have an initial
density field identical to the $128^3$ runs, constructed by sampling at
alternate grid points.

While our $4\hmpc$ box is small, well below the nonlinear scale at
$z=0$, we are interested here in the behavior on scales of a few kpcs,
and it is unlikely that the missing large-scale power would have a
significant effect on the inner portions of halos.  In addition, our
primarily conclusions are based on a {\it comparative} study between
collisionless and collisional dark matter for individual halos, so we
expect these results to be robust to volume effects.

\subsection{Modeling self-interactions}

\begin{deluxetable}{lc|cccc|cccc}
\footnotesize
\tablecaption{Simulation results.}
\tablewidth{0pt}
\tablehead{
\colhead{Model} &
\colhead{$\sdm (\xunits)$} &
\colhead{} &
\colhead{$128^3$\tablenotemark{\ast}} &
\colhead{} &
\colhead{} &
\colhead{} &
\colhead{$64^3$\tablenotemark{\dagger}} &
\colhead{} &
\colhead{}
}
\startdata
& & $N_{\rm halo}$ & $\alpha_{\rm med}$ & $\rho_{c,\rm med}$\tablenotemark{\ddag} & $c_{M,\rm med}$ & $N_{\rm halo}$ & $\alpha_{\rm med}$ & $\rho_{c,\rm med}$\tablenotemark{\ddag} & $c_{M,\rm med}$ \nl
\cline{3-10}
CDM  & 0          & 670 & $-1.49$ & 1.95 & 8.0 & 111 & $-1.58$ & 1.91 & 6.7 \nl
SIDM & $10^{-24}$ & 647 & $-0.93$ & 0.22 & 5.6 & 106 & $-1.44$ & 0.36 & 4.8 \nl
SIDM & $10^{-23}$ & 566 & $-0.37$ & 0.030 & 2.6 & 89  & $-0.53$ & 0.027 & 1.8 \nl
Observed & & & $\approx -0.5$ & $\approx 0.02$ & -- & & $\approx -0.5$ & $\approx 0.02$ & \nl
\enddata
\tablenotetext{\ast}{$128^3$ median values computed for all halos with $>1000$ particles ($\sim 30$ per run).}
\tablenotetext{\dagger}{$64^3$ median values computed for all halos with $>500$ particles ($\sim 10$ per run).}
\tablenotetext{\ddag}{Central dark matter density in $\Mpc3$; values shown 
are extrapolated from $1 h^{-1}\rightarrow 0.5$~kpc using $\alpha_{\rm median}$.}
\end{deluxetable}

We have modified GADGET to include self-interactions using a Monte Carlo
$N$-body technique to probabilistically incorporate collisions, along
the same lines as \cite{bur00} (2000) and \cite{koc00}, closer to the
latter as we use $\Delta {\bf v}$ from individual particles colliding
rather than setting $\Delta {\bf v}$ to be the particle's velocity; see
the discussion in \cite{koc00}.  Each pair of particles with positions and
velocities $({\bf r}_1,{\bf v}_1)$ and $({\bf r}_2,{\bf v}_2)$, separated
by $\delta x \equiv |{\bf r}_1-{\bf r}_2|/(2\epsilon)$ and $\delta v
\equiv |{\bf v}_1-{\bf v}_2|$, interact with a probability given by
\begin{equation}
P = f_{\rm geom}(\delta x)\; {\delta v\; \Delta t\over \lambda_X},
\end{equation}
where $\Delta t$ is the timestep, 
\begin{equation}
\lambda_X = {4\pi (2\epsilon)^3\over 3 m_p} {1\over \sdm}, 
\end{equation}
and
\begin{equation}
f_{\rm geom}(\delta x) = N {\int^1_0 W({\delta x}) W({\delta x}+{\delta x}') 
d({\delta x}')\over
{\int^1_0 W^2({\delta x}') d({\delta x}') } },
\end{equation}
where $W$ is the cubic spline kernel used in GADGET.  This geometrical
factor weights the probability of interaction by the product of spline
kernel-weighted density distributions of the two particles at their
given separation.  The normalization $N$ is set by requiring that
\begin{equation}
\int^1_0 f_{\rm geom}(\delta x)\; 4\pi\; \delta x^2\; d(\delta x) = 1,
\end{equation}
which ensures that when a particle has interacted with all its neighbors
within $2\epsilon$, the resulting probability is equivalent to
\begin{equation}
P = \sdm\;\rho\;\delta v\; \Delta t,
\end{equation}
where $\rho$ is the local dark matter density.

In our code, the scatterings are performed between individual particles at
the time that the acceleration between those particles is being computed
(i.e. during the ``treewalk").  In order to ensure that all possible
scatterings are considered, a tree cell is opened whenever it is within
$2\epsilon$ of a particle, regardless of the opening criterion.

If two particles scatter, their velocities are randomly re-oriented,
keeping the magnitudes of their velocities fixed.  In practice, a running
sum is kept of the change in velocity due to the interactions that a
given particle undergoes on every processor, and at the end of the step
the velocity change for each particle is summed over all processors and
added to that particle's velocity.  In this way, energy and momentum are
explicitly conserved, even if the scattered particles are on different
processors, or a particle undergoes more than one scatter in a single
timestep (which is very rare for the cross sections considered here).

We consider $\sdm=0$ (collisionless), $10^{-24}$, and $10^{-23}\xunits$.
We also examined $\sdm=10^{-25}\xunits$ and $\sdm=10^{-22}\xunits$ in
a $64^3$ simulation, which we will examine in \S\ref{sec: comparison}.
The total number of collisions per particle in our simulations are 1.01
for $\sdm=10^{-24}\xunits$, and 6.05 for $\sdm=10^{-23}\xunits$, with
slightly lower numbers (0.9 and 5.3) for the $64^3$ runs.  Note that
a factor of ten increase in $\sdm$ translates only to a factor of six
increase in the number of collisions, since the lowered central densities
(\S\ref{sec:  rhoc}) partially compensates for the increase in $\sdm$.
All runs were performed on Fluffy, a 32-processor Beowulf-class machine
at Princeton, with each $128^3$ run taking approximately one week.

\subsection{The Simulated Halo Sample}\label{sec: haloid}

We identify dark matter halos using SKID \footnote{\tt
http://www-hpcc.astro.washington.edu/tools/SKID/} (Spline Kernel
Interpolative DENMAX; see \cite{kat96}), with a linking length of
$2\epsilon$.  We only consider halos containing 64 or more particles,
to ensure a roughly complete sample of such halos in our simulations
(\cite{wei99}).  Table~1 lists the number of halos for identified in
these simulations.

A specific resolution issue arises from the finite number of particles
used to probabalistically model collisions in the SIDM simulations:
The number of particles in a given halo must be high enough to properly
Monte Carlo sample the distribution.  As we will show in \S\ref{sec:
res}, halos with $\ga 1000$ particles at $z=0$ seem to be accurately
represented with this technique for the simulations considered here.
This is quite restrictive, but still permits a significant sample of
halos (roughly 30 in each $128^3$ run) with which to compute statistics.
We also use the full sample of halos to examine certain aspects, but
we will be cautious about interpretations made from halos below this
``Monte Carlo resolution limit".

\section{Halo Structure}\label{sec: structure}

\subsection{Halo Profiles}\label{sec: prof}

We determine halo profiles $\rho(r)$ by spherical averages over radii
$r=\epsilon\rightarrow 30\epsilon$, in 20 equal intervals of $\log{r}$.
A sample of 16~halo profiles from our $128^3$ simulations is shown
in Figure~1.  Each panel shows a halo profile for $\sdm=0$ (solid),
$10^{-24}$ (short dashed), and $10^{-23}\xunits$ (long dashed).
Dotted line segments emanating from the innermost radius point of
the $\sdm=0$ curve indicate slopes $\alpha = -0.5, -1$ and $-1.5$
for comparison.  The same corresponding halos are chosen from each
simulation, allowing a case-by-case comparison of the effect of SIDM.
The outer halo profiles ($r\ga 10\hkpc$) are virtually identical for
each halo, showing that the effects of self-interactions are limited
to the inner few kpcs of halos, and confirming that the same halos are
being compared in the different simulations.  The halos in the leftmost
column are the four most massive ones in our simulations, while the halos
in other columns are chosen randomly from a descending range of masses.
Note that rightmost column shows halos with roughly 300 particles each,
below our Monte Carlo resolution requirement of $\ga 1000$ particles
(i.e. $\ga 3.6\times 10^9 M_\odot$), thus the effects of self-interactions
are not necessarily accurately represented in these cases.

From Figure~1 it is immediately evident that SIDM produces halos that have
enlarged central cores and shallower inner profiles.  CDM halos are almost
all cuspy ($\alpha\la -1$ typically), while most $\sdm=10^{-23}\xunits$
cores are close to flat.  $\sdm=10^{-24}\xunits$ leads to profiles that
are intermediate between these two.  In some cases, non-cuspy CDM
halos are seen, especially at lower masses.  In these cases the halo
may have undergone recent merging activity that temporarily lowers the
central density, which is particularly effective in smaller mass halos.
Additionally, recent mergers that have not relaxed make it difficult to
unambiguously identify the halo center about which to compute profiles,
typically making profiles appear shallower.  We make no cut in regards
to the merging history or ``isolatedness" of halos, but we do note that
the missing large-scale power in our simulations will tend to generate
fewer mergers, and make the largest objects in our simulations appear more
isolated.

\PSbox{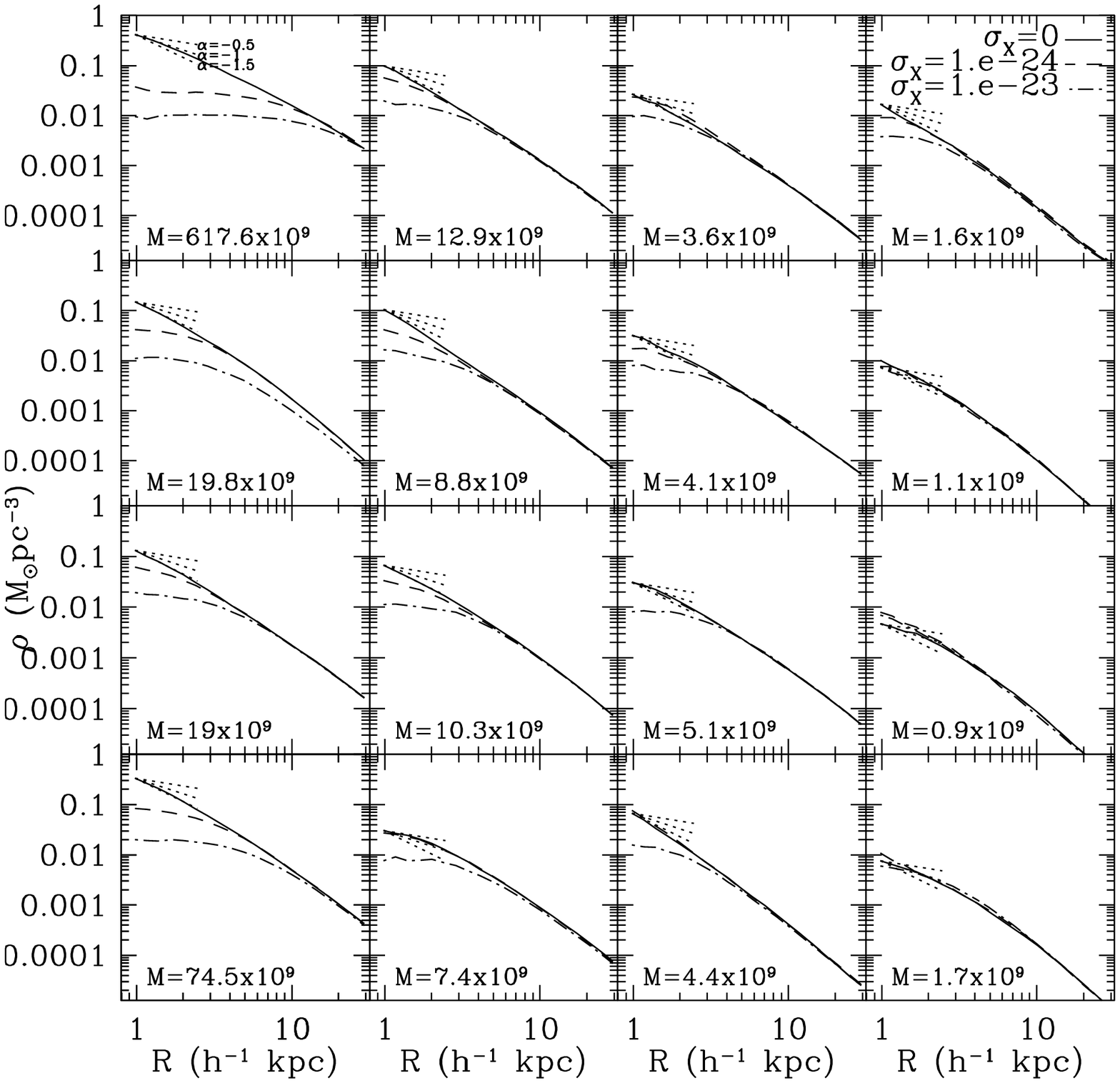  angle=0 voffset=-70 hoffset=-30 
vscale=48 hscale=48}{3.0in}{4.0in} 
{\\\small Figure~1: 16 selected halo profiles for collisionless
(solid), $10^{-24}\xunits$ (dashed), and $10^{-23}\xunits$ (dot-dashed).
Corresponding halos from each simulation are presented, allowing a direct
comparison of the effect of SIDM on a halo-by-halo basis.  The total halo
mass in $M_\odot$ for the $\sdm=0$ halo is shown in the lower right; the SIDM halo
masses are typically within 20\%.  The columns are ordered by mass, with
the four highest mass halos shown in the leftmost column.  Dotted lines
from innermost point show reference slopes of $\alpha=-0.5,-1.0,-1.5$.
\vskip0.1in
}

At high masses, the effect of SIDM is very prominent.  The upper
leftmost halo is Milky Way-sized ($6\times 10^{11}M_\odot$), and shows
a large core of $\sim 15\hkpc$ for $\sdm=10^{-23}\xunits$ ($\sim
8\hkpc$ for $\sdm=10^{-24}\xunits$).  The core size and difference in
inner slope become less prominent to lower masses, though this could be
due to the increasing effects of unrelaxed halos, as well as the Monte
Carlo resolution issues discussed earlier.  We examine these issues
quantitatively in \S\ref{sec: alphamass}.

No evidence is seen for SIDM halo profiles that are isothermal,
as would be expected if the cross section was so large that core collapse
would occur on timescales significantly shorter than a Hubble time.
This supports the analytic estimates of \cite{spe00} that core collapse
on a Hubble time would not occur until $\sdm\ga 10^{-22}\xunits$.

Overall, there is a clear trend on a case-by-case basis that SIDM
results in a reduced central density and shallower inner slope of the
dark matter halo, with increasing $\sdm$ having a greater such effect.

\subsection{Central Densities}\label{sec: rhoc}

Figure~2 shows the central density of dark matter halos $\rho_c$, taken
to be the density at our innermost resolved radius $\epsilon$, as a
function of halo mass.  Here we only consider halos with more than 1000
particles, where our Monte Carlo technique has sufficient numbers to
represent the collisional behavior (as we will discuss in \S\ref{sec:
res}).  The central halo density of galaxies is observed to be $\sim
0.02 \Mpc3$ (\cite{fir00b}), and is consistent with being independent
of halo mass.  The observed range of halo densities is shown as the
hatched region, with a majority of their data falling towards the lower
end of that region.  The arrow in the upper left indicates the increase
in $\rho_c$ projecting the profile from $1\hkpc$ in to 500~pc, typical
of observations of dwarf and low surface brightness galaxy central
densities, using the slope shown.

SIDM $\sdm=10^{-23}\xunits$ halos are in good agreement with these
observations, while $\sdm=10^{-24}\xunits$ produces inner densities that
are a few times higher, but still marginally consistent with observations.
In addition, $\rho_c$ in SIDM models show little trend with halo mass,
in agreement with observations, because the core density is set by
collisional physics.

\PSbox{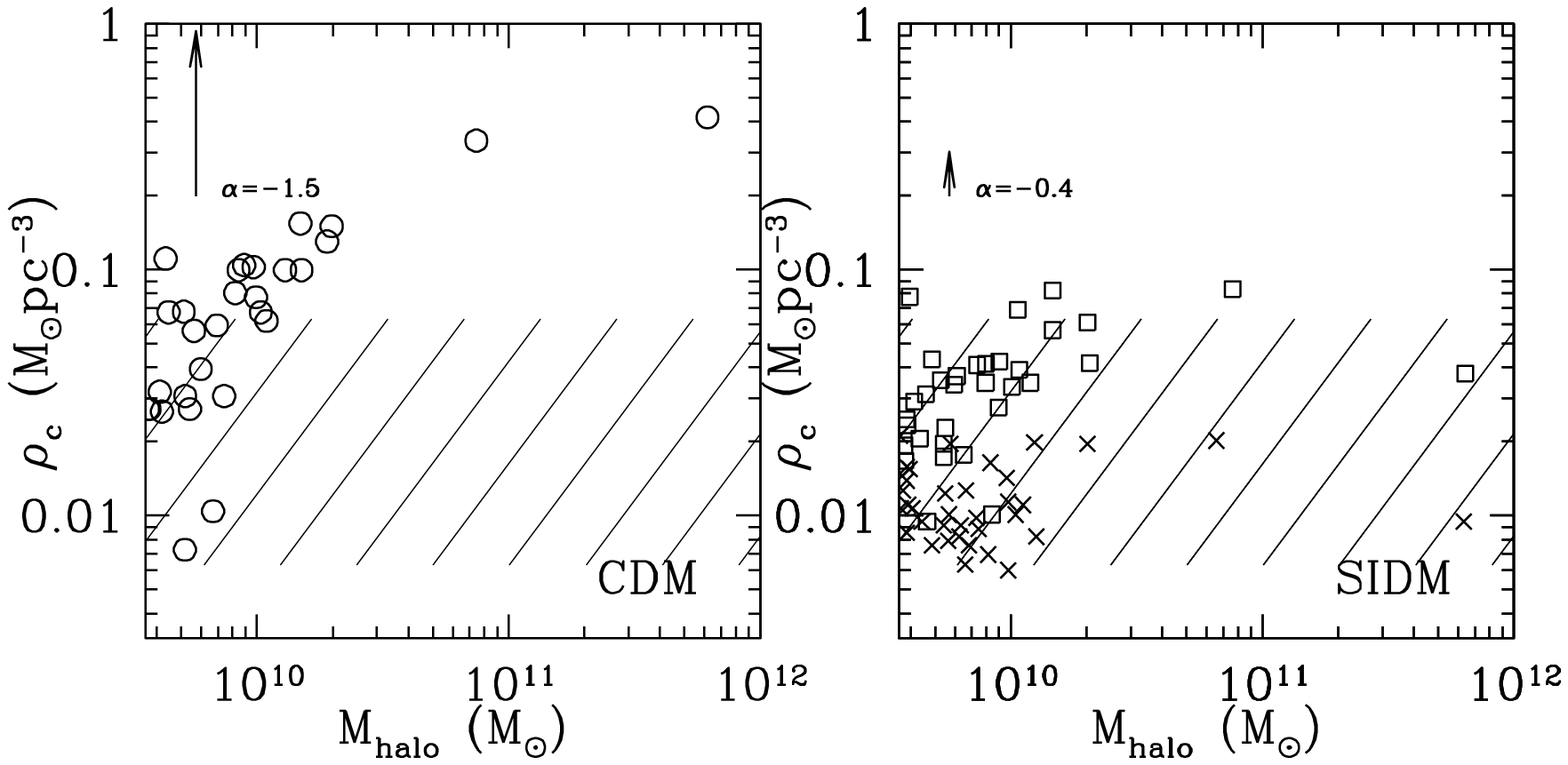 angle=0 voffset=-220 hoffset=-40
vscale=48 hscale=48}{3.0in}{2.0in} 
{\\\small Figure~2: $\rho_c$ vs. $M_{\rm halo}$ for CDM (left panel)
and SIDM (right panel).  For SIDM, crosses show $\sdm=10^{-23}\xunits$,
while open squares show $\sdm=10^{-24}\xunits$.  Only halos with 1000
or more particles are shown.  The hatched region indicates the range of
observed $\rho_c$ compiled by \cite{fir00a}.  Arrow in upper left
indicates how much each value of $\rho_c$ would increase if measured at
500~pc (instead of $1\hkpc$), typical of observations, assuming a
profile with the slope shown.
\vskip0.1in
}

Conversely, the more massive halos in CDM have central densities that
are too high by at least an order of magnitude already at $1\hkpc$, and
because of their cuspy profile the disagreement would be much worse at
smaller radii, as indicated by the arrow in the upper left.  Moreover,
CDM halos have central densities that increase with mass, in conflict
with observations.

As such, SIDM halos appear to agree better with observations.  Table~1
lists the median central density of halos with more than 1000 particles
in our various models.  This shows that our simulations reproduce the
observed central halo densities for $\sdm\approx 10^{-23}\xunits$.

\subsection{Inner Profile Slopes}\label{sec: alpha}

We estimate the inner halo profile slope $\alpha$ as the slope between the
innermost resolved radii, $r=1\rightarrow 1.5\hkpc$.  Figure~3 shows a
histogram of this slope for the collisionless (solid line),
$10^{-24}\xunits$ (dashed), and $10^{-23}\xunits$ (dot-dashed) cases,
for all halos that have more than 1000 particles.  The number of such
halos in each simulation is indicated in the legend.  The qualitative
impression from Figure~1 that SIDM produces shallower inner profiles is
quantified in Figure~3.  The median values of $\alpha$ are indicated by
the arrows from the upper $x$-axis, and are listed in Table~1.

CDM produces halos that have cuspy cores, with $\alpha_{\rm med}\approx
-1.5$.  This is consistent with the work of \cite{moo99}, among
others.  25 of the 28 CDM halos have $\alpha<-1$, indicating that cuspy
cores are a common feature of CDM models.  Conversely, the inner slopes
in SIDM models are significantly shallower.  For
$\sdm=10^{-24}\xunits$, $\alpha_{\rm med}\approx -0.9$, while for
$\sdm=10^{-23}\xunits$, $\alpha_{\rm med}\approx -0.4$, with no halos
having $\alpha<-1$.  This latter case has a median $\alpha$ close to
the value preliminarily suggested by H$\alpha$ observations of low
surface brightness galaxies (\cite{dal00}), though a definitive value
awaits a more thorough analysis of observational biases.

\PSbox{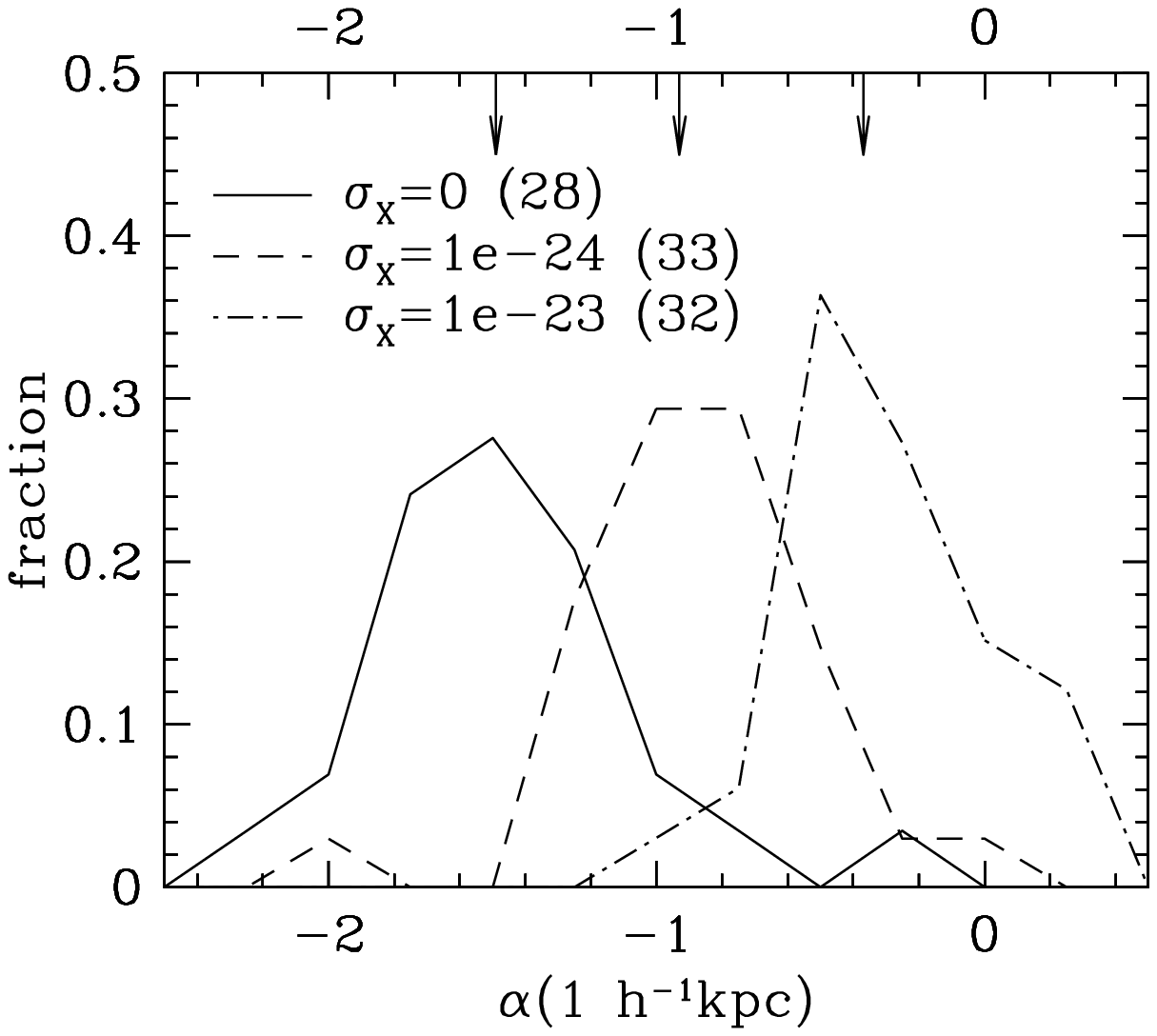 angle=0 voffset=-200 hoffset=-30
vscale=48 hscale=48}{3.0in}{2.3in} 
{\\\small Figure~3: Histogram of inner slopes $\alpha$
for collisionless (solid), $10^{-24}\xunits$ (dashed), and
$10^{-23}\xunits$ (dot-dashed).  Only halos with $M>3.6\times 10^9
M_\odot$ ($>1000$ particles) are included; the number of such halos is
indicated in parantheses in the legend.  The median values of $\alpha$
are indicated by the arrows from the top edge of the plot, and are listed
in Table~1.
\vskip0.1in
}

The scatter in $\alpha$ is mostly real.  There is some scatter due to
discreteness effects in measuring the inner slope as we do.  While we
could obtain an inner slope from fitting a general halo shape
(\cite{her90}; \cite{kly00}), with 5 free parameters the inner slope
would be poorly constrained by 20 correlated data points, thus we
choose our simpler definition.  Further scatter arises from recently
merged halos that temporarily have shallower profiles until relaxed.
However, neither of these effects is very significant for the large
mass halos plotted in Figure~3.  Still, we choose to quote the median
$\alpha$ rather than the mean, in order to quantify ``typical" halos in
these models and reduce sensitivity to outliers, although the mean is
similar.  We note that a significant scatter in inner slopes is also
seen in the observations (e.g. \cite{deb96}; \cite{dal00}).

Our CDM profiles are, at face value, in better agreement with the
analytic profile of \cite{moo99} (1999), with an asymptotic slope of
$\alpha=-1.5$, rather than an \cite{nfw96} profile having
$\alpha(r\rightarrow 0)=-1$.  However, profile fitting is a tricky
business (as discussed in \cite{kly00}).  By reducing the scale
radius in the \cite{nfw96} profile (i.e. increasing the concentration),
one can push the transition to a slope of $\alpha=-1$ to a radius
smaller than $1\hkpc$ where we cannot resolve the profiles (the
``cusp-core degeneracy"; see \cite{vs00}).  Thus we
suspect that our CDM profiles can also be adequately fit by an
\cite{nfw96} profile having a large concentration parameter.  As such,
we do not argue for or against either profile form.  Our simulations
can only predict the slope at $r\approx 1\hkpc$, and that is what
should be compared to observations.

SIDM appears to be in better agreement with observations of the inner
slopes of dark halo profiles than CDM.  At face value,  $\sdm=
10^{-23}\xunits$ is preferred, but given uncertainties in observations
and simulation techniques, $\sdm= 10^{-24}\xunits$ is probably also
consistent.  A similar value of $\sdm$ also reproduces the observed
central density of galaxies.  Such a coincidence is not expected {\it a
priori}, and may represent a significant success of the SIDM scenario.

\subsection{Mass Dependence of Inner Slope}\label{sec: alphamass}

Figure~4 shows a plot of inner halo slope $\alpha$ vs. halo mass $M_{\rm
halo}$ for all halos in our CDM (left panel) and SIDM (right panel,
$\sdm=10^{-23}\xunits$) simulations.  The curve shows the running median
value of $\alpha$ in bins of $\Delta (\log{M}) = 0.5$.

\PSbox{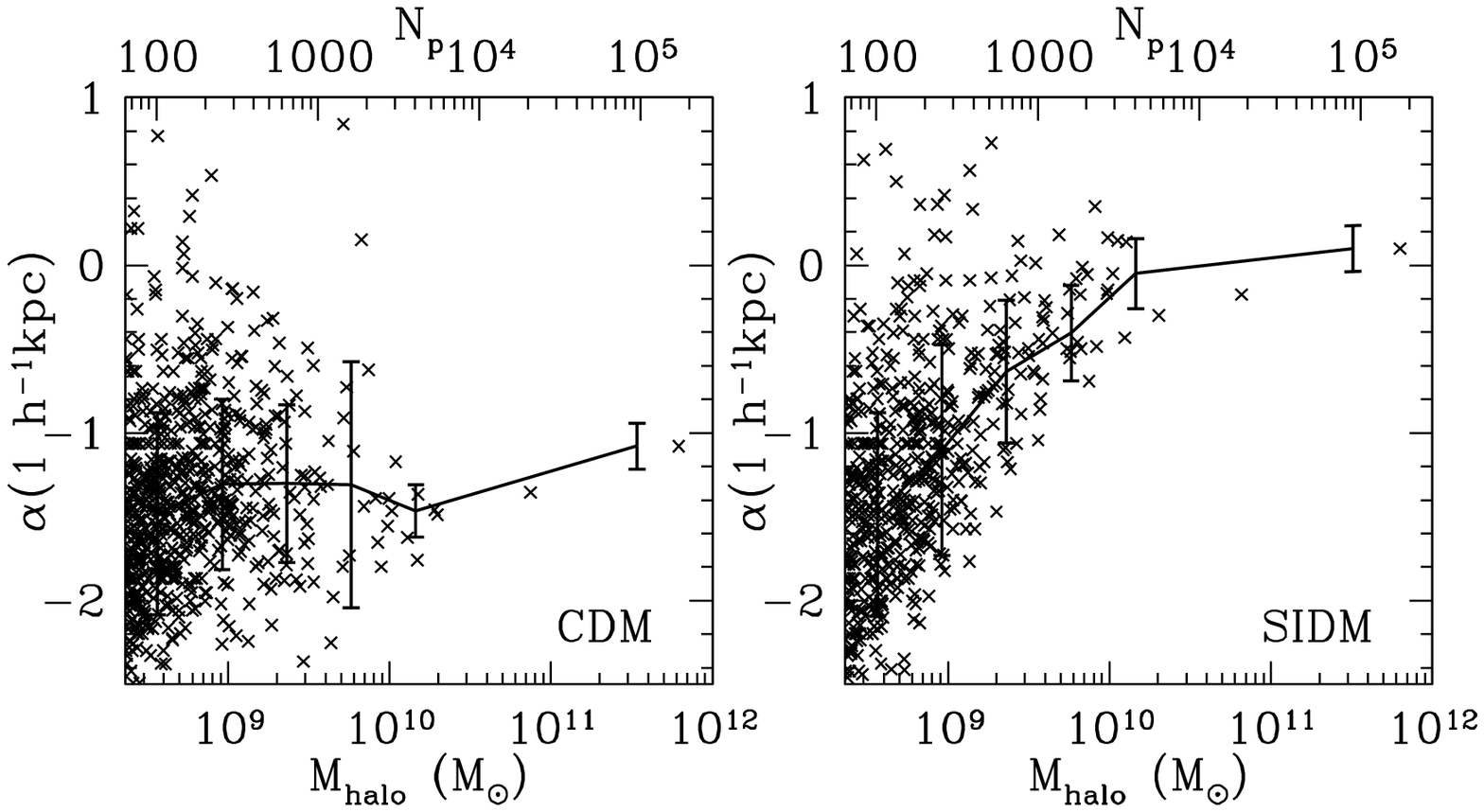 angle=0 voffset=-220 hoffset=-40
vscale=48 hscale=48}{3.0in}{2.0in} 
{\\\small Figure~4: $\alpha$ vs. $M_{\rm halo}$
for CDM (left panel) and SIDM ($\sdm=10^{-23}\xunits$; right panel).
All halos with 64 or more particles are shown.  The line shows a running
median of the $\alpha$ distribution, with a variance computed from
all halos within each mass bin.
\vskip0.1in
}

The CDM case shows almost no trend with mass, with the median slope
always around $-1.2 \sim -1.5$.  The largest halo does have a slightly
shallower slope, consistent with the trend seen in very high resolution
CDM simulations of individual halos (\cite{jin00}).  The scatter increases
to low mass due to discreteness and merging effects described in
\S\ref{sec: alpha}.

The SIDM case shows dramatically different behavior, suggesting at face
value that smaller halos have steeper inner profiles.  However, caution
must be used in interpreting this result.  First, smaller halos have
smaller cores, meaning that a slope measured at a fixed radius (not
scaled to the halo core size) will result in a steeper slope.  For CDM
this effect is less significant, since the slope remains similar from
the outer to the inner halo.  Second, the small numbers of particles in
the low-mass halos makes the Monte Carlo technique less effective in
modeling collisions, thereby making SIDM appear more like CDM; we
investigate this issue further in \S\ref{sec:  res}.  Thus we make
no claim regarding a trend of $\alpha$ with $M_{\rm halo}$.

\subsection{Mass Concentration Parameter}\label{sec: conc}

As seen in Figure~1, SIDM appears to have the desired effect of reducing
the concentration of dark matter halos.  In this section we quantify
this effect using a concentration parameter, which we define differently
than previous authors in order to facilitate a more direct comparison with
observations.

The canonical definition of a concentration parameter is given by
\cite{nfw96} as the ratio between the virial radius $\r200$ (taken to
be the radius at which the halo density is $200\times$ the cosmic mean)
and the scale radius of the halo $r_s$ in the \cite{nfw96} profile.
This concentration parameter, however, is difficult to compute
unambiguously in the case of non-isolated halos, and difficult to
compare directly to observations that seldom extend out to $\r200$.
Furthermore, $r_s$ is only defined within the context of the specific
\cite{nfw96} model, and profile fits are typically degenerate between
$r_s$ and concentration (\cite{kly00}).  Col\'in et al. (2000) circumvent some
of these issues by defining the concentration as the ratio of the
minimum of $\r200$ and the halo radius to the radius that encloses
20\% of the halo mass.  However, this inner radius is dependent on
knowing the total halo mass, something which is difficult to determine
observationally.

Instead, we choose to define a {\it mass concentration parameter} $c_M$, based
on enclosed mass rather than radii, and restrict the scales in our
definition to those where observations are available, typically $r\la
20\hkpc$.  We define
\begin{equation}\label{eqn: cm}
c_M = 27 {M(<r_{\rm in})\over M(<r_{\rm out})},
\end{equation}
with 
\begin{equation}\label{eqn: rin}
r_{\rm in} = \frac{1}{3} r_{\rm out} = 8.5\; {\rm kpc}\; {v_{\rm circ}\over{220 \kms}}\;,
\end{equation}
where $v_{\rm circ}$ is the circular velocity of the halo.  The choice
of $r_{\rm in}$ is arbitrary; here we base it on the Milky Way, as it
is convenient and results in observationally accessible scales.  The
scaling with $v_{\rm circ}$ is that expected for self-similar halos
following the Tully-Fisher relation.  The normalization factor of 27
results in a uniform density distribution having a mass concentration
of unity.  A flat rotation curve between $r_{\rm in}$ and
$r_{\rm out}$ implies $M(r)\propto r$, resulting in $c_M=9$.  In our
simulations, we take $v_{\rm circ}$ to be the maximum circular velocity
of the halo as output by SKID.

\PSbox{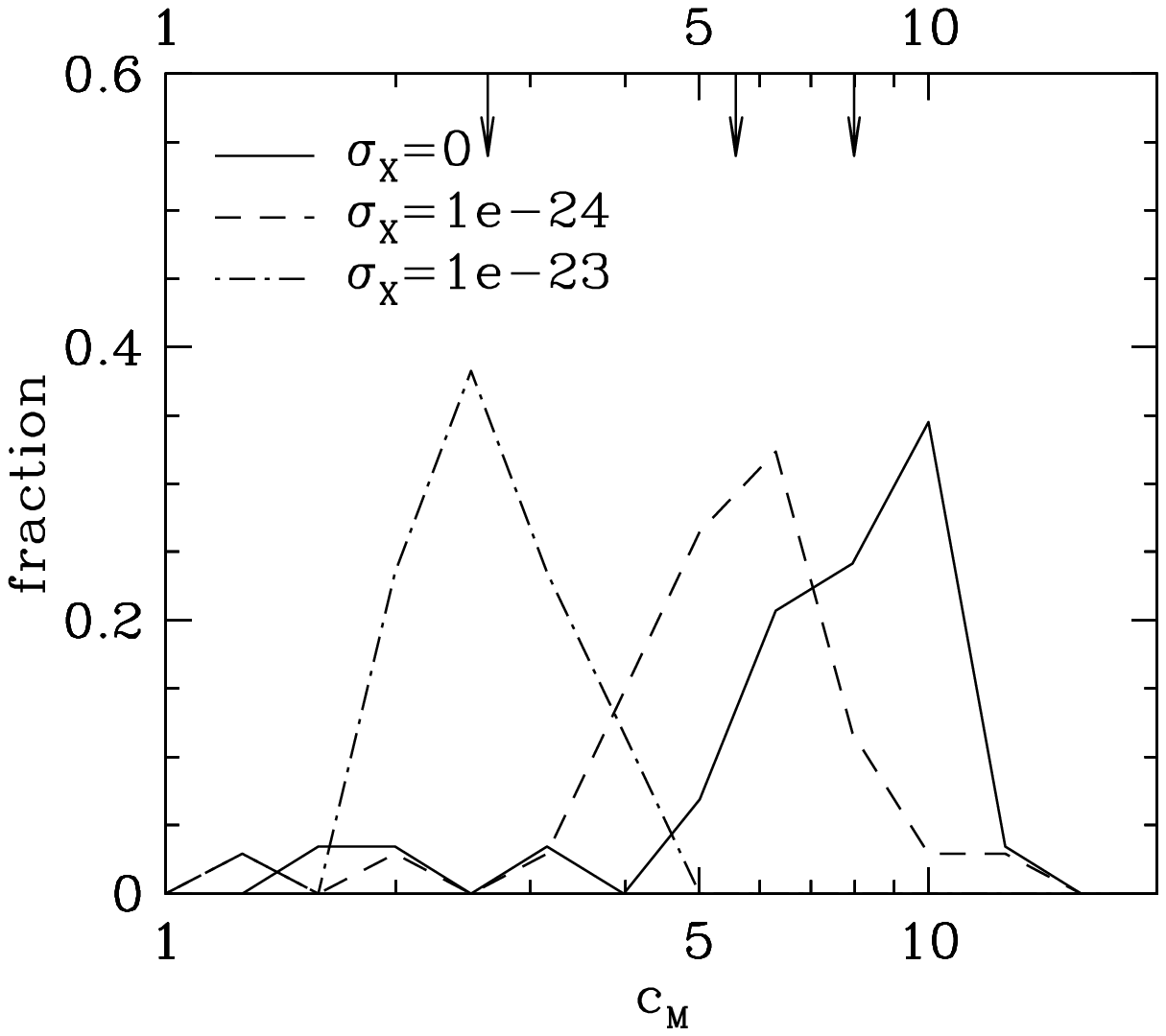 angle=0 voffset=-200 hoffset=-30
vscale=48 hscale=48}{3.0in}{2.3in} 
{\\\small Figure~5: Histogram of mass concentrations $c_M$ for
collisionless (solid), $10^{-24}\xunits$ (dashed), and
$10^{-23}\xunits$ (dot-dashed).  Only halos with $M>3.6\times 10^9
M_\odot$ ($>1000$ particles) are included.  The median values of
$c_M$ are indicated by the arrows from the top edge of the
plot, and are listed in Table~1.
\vskip0.1in
}

Figure~5 shows a histogram of $c_M$ for all halos with more than 1000
particles.  As expected, there is a clear trend for CDM to have more
concentrated halos than SIDM, with the amount of concentration
decreasing with increasing $\sdm$.  Note that the difference between
SIDM with $10^{-24}\xunits$ and $10^{-23}\xunits$ is exaggerated
relative to the difference between the inner slopes of those models
(cf. Figure~3).  This is because the concentration is increased in
$10^{-24}\xunits$ relative to $10^{-23}\xunits$ both due to the
increased inner slope, as well as the reduced core radius.  We also
examined the mass dependence of $c_M$ and found no obvious trend,
but our range of masses is small.

For comparative purposes, our mass concentration parameter $c_M$ may be
analytically related to the \cite{nfw96} concentration parameter $\cnfw$.
From \cite{nfw96},
\begin{equation}\label{eqn: vc}
{v_{\rm c}^2(x)\over \v200^2} = \frac{1}{x} {\ln{(1+\cnfw x)}-\cnfw x/(1+\cnfw x)\over {\ln{(1+\cnfw)}-\cnfw/(1+\cnfw)} },
\end{equation}
where $x=r/\r200$, $v_{\rm c}(x)$ is the circular velocity at $x$, and
$\v200$ is the circular velocity at $\r200$.  Our $v_{\rm circ}$ is taken to
be the maximum halo circular velocity, which may obtained by
maximizing equation~\ref{eqn: vc}; this occurs at $x_{\rm max}\approx
2/\cnfw$ (though we compute it exactly for the results shown below).

Let $\hat{v}_{\rm circ}\equiv v_{\rm c}(x_{\rm max})/\v200$.  In 
appropriate units, $\v200=\r200$ (see \cite{nfw96}, equation~A2).
Thus
\begin{equation}
x_{\rm in}\equiv r_{\rm in}/\r200 = \frac{8.5}{220} \hat{v}_{\rm circ},
\end{equation}
implying $x_{\rm in}$ and $x_{\rm out}=3x_{\rm in}$ are solely
functions of $\cnfw$ (note that this arises because we defined 
$r_{\rm in}\propto v_{\rm circ}$).  Using $M(<r)\propto r v_{\rm c}^2(r)$,
\begin{equation}
c_M = 9 \frac{v^2_c(x_{\rm in})}{v^2_c(x_{\rm out})},
\end{equation}
which is purely a function of $\cnfw$.
The resulting relationship is shown in Figure~6.

\PSbox{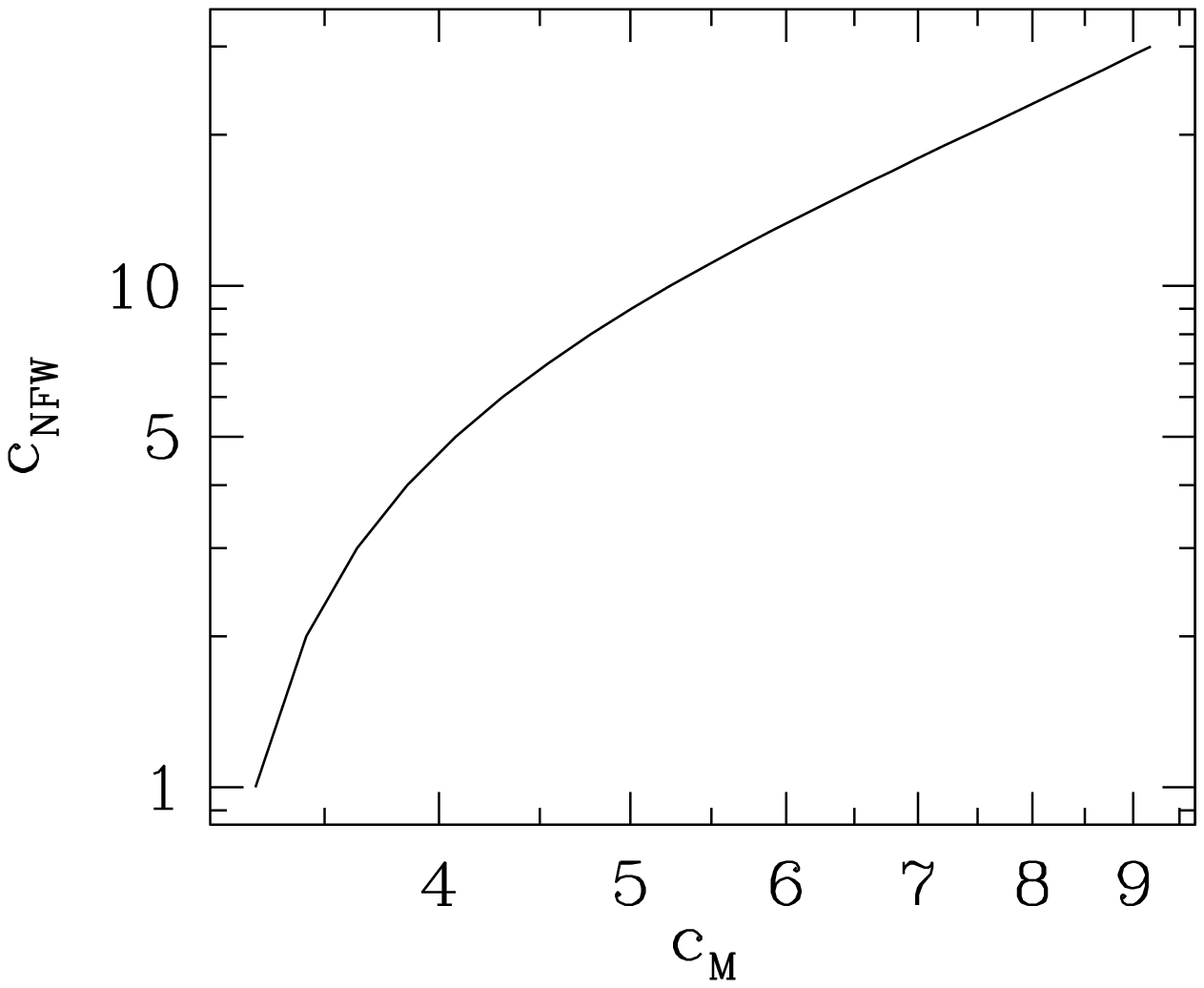 angle=0 voffset=-200 hoffset=-30
vscale=48 hscale=48}{3.0in}{2.3in} 
{\\\small Figure~6: Mass concentration parameter $c_M$, defined
in equation~\ref{eqn: cm}, vs. \cite{nfw96} concentration parameter
$\cnfw$.
\vskip0.1in
}

Figure~6 shows that $c_M\approx 8$, typical of halos in our CDM model,
corresponds to $\cnfw\approx 23$.  This value is in agreement with
expectations for dwarf galaxies in a $\Lambda$CDM model.  Conversely,
$c_M\approx 5.6$, which is the median value for SIDM with
$10^{-24}\xunits$, corresponds to $\cnfw\approx 11$.  Note that the
minimum value of $c_M$ for an \cite{nfw96} halo is 3.  Thus SIDM with
$10^{-23}\xunits$, with $c_{M, \rm med}\sim 2.6$, produces halos that typically
cannot be described properly by $\cnfw$.  This is because these halos
have $\alpha>-1$ typically, so \cite{nfw96} profiles with
$\alpha(r\rightarrow 0)=-1$ are a poor fit.  This further illustrates
why $\cnfw$ is a poor way to describe halos in general.

The largest halo in our simulations has a mass comparable to the Milky
Way's, $\approx 6\times 10^{11} M_\odot$.  The concentrations of this
halo are 6.7, 4.4 and 2.3 in CDM, $\sdm=10^{-24}\xunits$ and
$\sdm=10^{-23}\xunits$, respectively.  $c_M$ of the Milky Way halo is
somewhat uncertain, because of the uncertainty in the rotation curve
outside the solar circle ($R_\odot$) and the effect of baryonic mass
within $R_\odot$, but we make a rough estimate here.  If the rotation
curve is flat, $c_M=9$ as stated before.  There are suggestions that
the rotation curve rises somewhat beyond the solar circle (though this
is uncertain; see \cite{oll00}), in which case $c_M$ is reduced; let us
take $c_M=8$ as a working estimate.  The rotation curve measures the
total mass, so we must correct for the baryons to compare with our
simulated $c_M$.  If we take the fraction of baryonic mass to be 50\%
inside $R_\odot$ and 20\% inside $3R_\odot$, then $c_M$ reduces to 5.
In addition, baryons adiabatically compress the dark matter as they
dissipate, so we must correct the Milky Way $c_M$ further downwards to
compare to our dissipationless halos.  From the analysis of \cite{avi98},
this reduction factor is $\sim 1.5-2$, resulting in the Milky Way halo
having $c_M\sim 3$.  Thus after reasonable corrections, the Milky Way
mass concentration appears to be in better agreement with SIDM than
CDM.  The rapid rotation of bars also suggests a lower concentration
for the Milky Way-sized galaxies than that predicted by CDM
(\cite{deb98}).

A more direct comparison with simulations may be obtained from rotation
curves of dark matter-dominated, low surface brightness galaxies, where
baryonic corrections are smaller.  We expect that this mass
concentration measure $c_M$ will be relatively straightforward to
compute from such rotation curves (e.g. \cite{dal00}), so we look
forward to comparisons.  $c_M$ has the advantage that it is independent
of halo fitting parameters, as the enclosed mass can be obtained
directly from the observed circular velocity with modest assumptions.
In this sense, it is a more robust comparison than the inner slope and
the \cite{nfw96} concentration parameter, which are degenerate and
sensitive to scales outside those typically observed (\cite{vs00}), and
the central density, which depends on an uncertain contribution from
baryons.

\subsection{Phase Space Densities}\label{sec: spaze face}

A recently popularized measure of the concentration of dark matter
halos is the central phase space density.  \cite{dal00a} find that
observed phase space densities $Q\equiv \rho/\sigma^3$ scale as
$Q\propto \sigma^{-3}-\sigma^{-4}$, where $\sigma$ is the velocity
dispersion, from dwarf spheroidals up to clusters of galaxies.
Observations compiled by \cite{sel00} suggest a similar relation,
albeit with a large scatter, and he uses them to argue against {\it
any} form of collisionless dark matter (though see \cite{mad00}).

Figure~7 shows the phase space density $Q$ of dark matter within
$r_{\rm in}$ as a function of $\sigma$, for halos with
$\sigma>30\kms$.  We calculate $\sigma$ as the velocity dispersion
around the group center of mass velocity, within $r_{\rm in}$ 
(cf. equation~\ref{eqn: rin}).  Open circles show CDM halos, crosses
indicate SIDM ($\sdm=10^{-23}\xunits$) halos.  The dashed and dotted
lines show $Q\propto \sigma^{-3}$ and $Q\propto \sigma^{-4}$,
respectively, that bracket the observations, reproduced from
\cite{dal00a}.

\PSbox{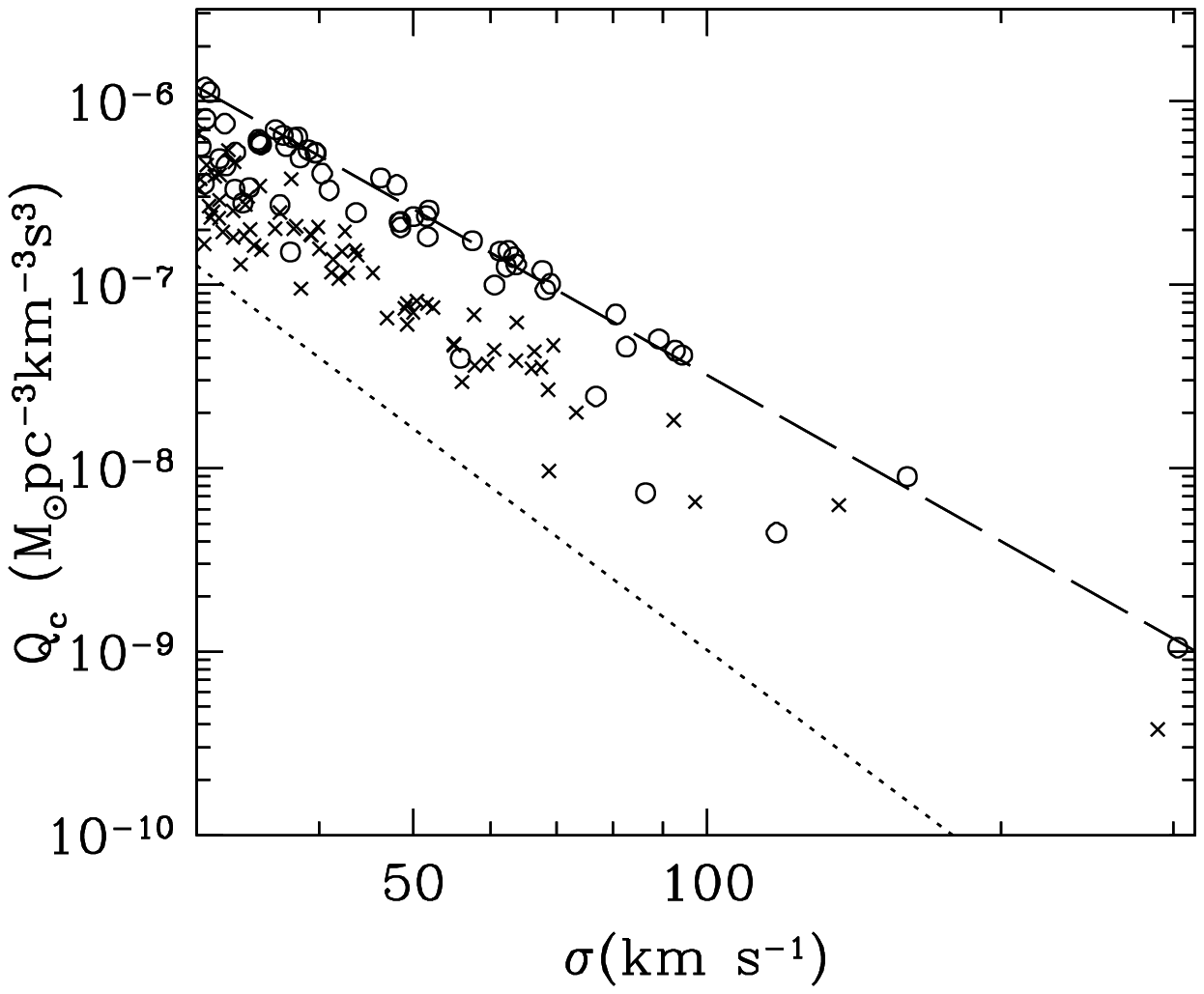 angle=0 voffset=-200 hoffset=-30
vscale=48 hscale=48}{3.0in}{2.3in} 
{\\\small Figure~7: Phase space density $Q$ vs. velocity dispersion
$\sigma$ for CDM (open circles), and SIDM with $\sdm=10^{-23}\xunits$
(crosses).  Dashed and dotted lines bracket observations, showing
scalings of $Q\propto \sigma^{-3}$ and $Q\propto \sigma^{-4}$,
respectively.
\vskip0.1in
}

SIDM generally shows lower phase space densities than CDM.  SIDM is in
somewhat better agreement with observations, falling in the middle of
the observed range.  Measuring $Q$ in galaxies is a difficult task,
because the stellar velocity dispersion is not necessarily that of the
dark matter.  Furthermore, in rotationally supported galaxies the
dubious assumption of an isothermal spherical halo is used to relate
circular velocity to dispersion.  Thus $Q$ is perhaps not among the
most useful observational discriminants between CDM and SIDM.

An interesting remark from Figure~7 is that the scaling of $Q(\sigma)$
is roughly the same in both models, roughly $Q\propto \sigma^{-3}$.
\cite{dal00a} argue that such a scaling results from the dynamical
assembly of halos, and is not expected based on simple phase packing
arguments.  This further motivates simulations of SIDM that include the
cosmological growth of halos via dynamical processes of merging and
accretion.

\subsection{Ellipticities}\label{sec: ellipt}

SIDM produces halos that are more spherical than CDM, because of the
isotropic nature of the collisions (\cite{spe00}).  This is a generic
feature of SIDM with {\it any} significant cross section, since in the
inner portions of halos where collisions are frequent, the velocity
ellipsoid is quickly isotropized.  Thus the shapes of dark matter halos
provide an important observational discriminant between CDM and SIDM.

We compute axis ratios of our halos using the prescription outlined
in \cite{dub91}.  They define a tensor
\begin{equation}
M_{ij} = \sum {x_i x_j\over a^2},\;\;\; {\rm with}\;\;\; 
a\equiv \biggl(x_1^2+{x_2^2\over q^2}+{x_3^2\over s^2}\biggr)^\frac{1}{2},
\end{equation}
where $a$ is the elliptical radius, and $s\leq q\leq 1$ are the axis
ratios, and the sum is over all particles with distances
$\approx (x_1,x_2,x_3)$ from the halo center along the axes of the
ellipsoid.  Then,
\begin{equation}
q = \biggl({M_{yy}\over M_{xx}}\biggr)^\frac{1}{2}\;\;\; {\rm and}\;\;\; s = \biggl({M_{zz}\over M_{xx}}\biggr)^\frac{1}{2}\;,
\end{equation}
where $M_{xx}\geq M_{yy}\geq M_{zz}$ are the eigenvalues of $M$.  $q$
represents the axisymmetry of the halo, while $s$ measures the halo
flattening.  Since $a$ depends on $q$ and $s$, the calculation of
$M_{ij}$ must be iterated until convergence, which we take to be better
than 0.01 in $q$ and $s$.  This scheme weights particles roughly
equally regardless of distance from center, unlike a moment of inertia
tensor which weights the outskirts heavily, and thus better represents
the ellipticity of the density distribution, as shown in \cite{dub91}.

\PSbox{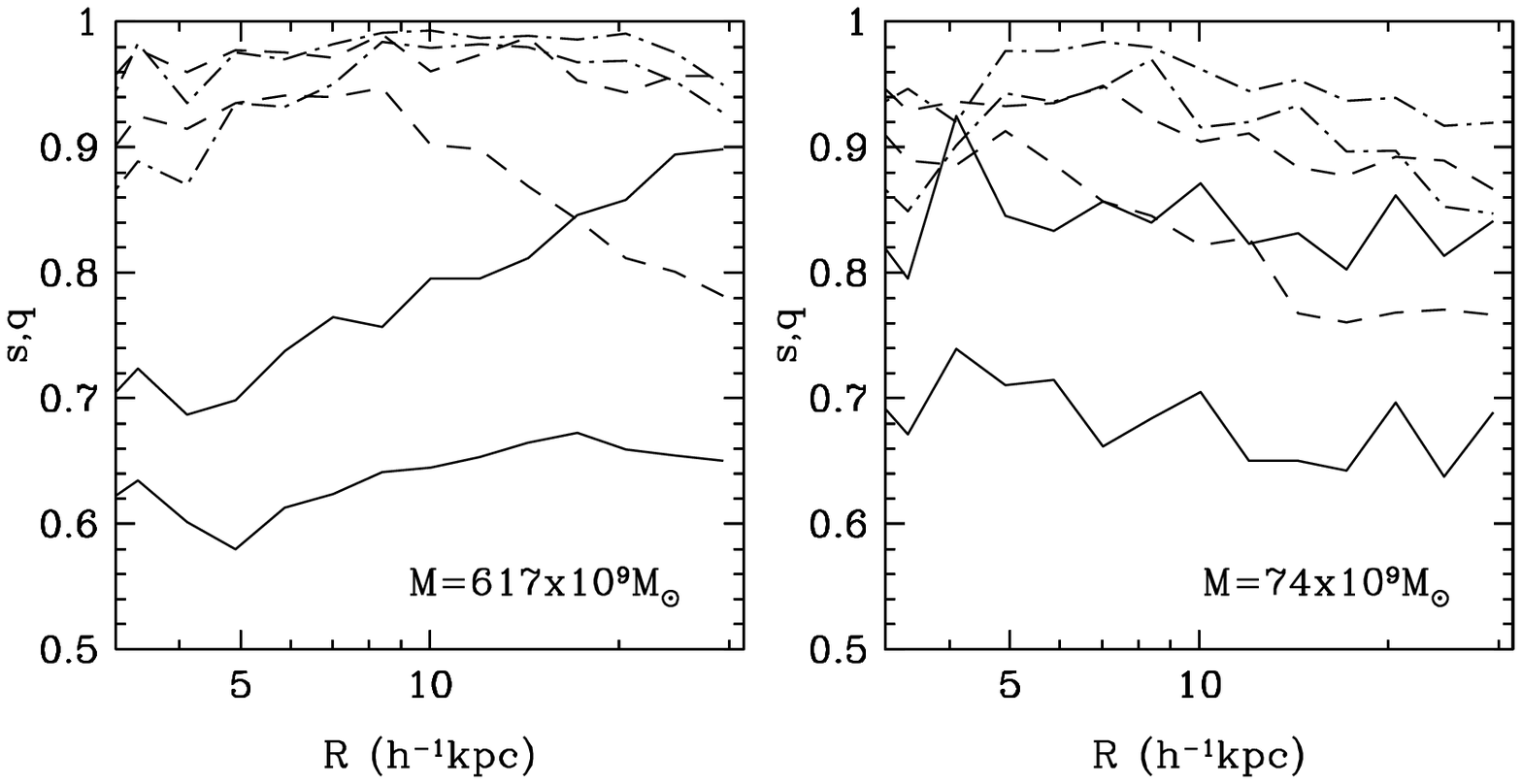 angle=0 voffset=-220 hoffset=-40
vscale=48 hscale=48}{3.0in}{2.0in} 
{\\\small Figure~8: Axisymmetry $q$ and flattening $s$ vs. $r$ for a
$6\times 10^{11} M_\odot$ halo (left panel) and a $7\times 10^{10}
M_\odot$ halo (right panel).  Solid lines are CDM, dashed lines are
SIDM, $\sdm=10^{-24}\xunits$, and dot-dashed lines are SIDM,
$\sdm=10^{-23}\xunits$.  Lower of two curves is $s$.
\vskip0.1in
}

Figure~8 shows axis ratios as a function of radius in our two most
massive halos, having masses $\approx 6\times 10^{11} M_\odot$ (left
panel) and $\approx 7\times 10^{10} M_\odot$ (right panel).  Solid line
is the CDM halo, dashed line is SIDM with $\sdm=10^{-24}\xunits$,
and dot-dashed line is SIDM with $\sdm=10^{-23}\xunits$.  $q$ is the
upper of two curves for a given model.  

CDM halos are fairly triaxial, while SIDM produces halos that are
much closer to spherical.  The effect is dependent on radius, as in
the outer regions SIDM and CDM become more similar, since the effect
of self-interactions is confined to the inner parts of halos.  Still,
even at $30\hkpc$ there are significant differences between SIDM and CDM.
While not stated, this trend with radius is also evident from Figure~1
of \cite{yos00b}.

Figure~9 shows histograms of axis ratios $q$ and $s$ at $2\hkpc$ (top
panels) and $10\hkpc$ (bottom panels) for all halos with more than
1000 particles.  The median value for each model is indicated by the
corresponding tick mark on the top axis.  The difference between CDM and
SIDM is more pronounced at small radii, where CDM produces significantly
triaxial halos while SIDM halos remain spherical.  At large radii there
is a much milder trend to more spherical halos with increasing $\sdm$.
$s$ also shows more differences than $q$.

Figure~9 shows that while CDM produces halos are typically more
spherical, there is still significant non-sphericity in many SIDM halos.
In particular, there is a tail in the distributions of both CDM and
SIDM to smaller axis ratios.  This may be due to asymmetric infall
that temporarily distorts the shape of the density in some halos,
particularly smaller ones.  This also may just be an artifact of finite
number of collisions in smaller halos.  Note that the two largest halos
shown in Figure~8 show greater differences at $10\hkpc$ than suggested
by the statistics in Figure~9.

\PSbox{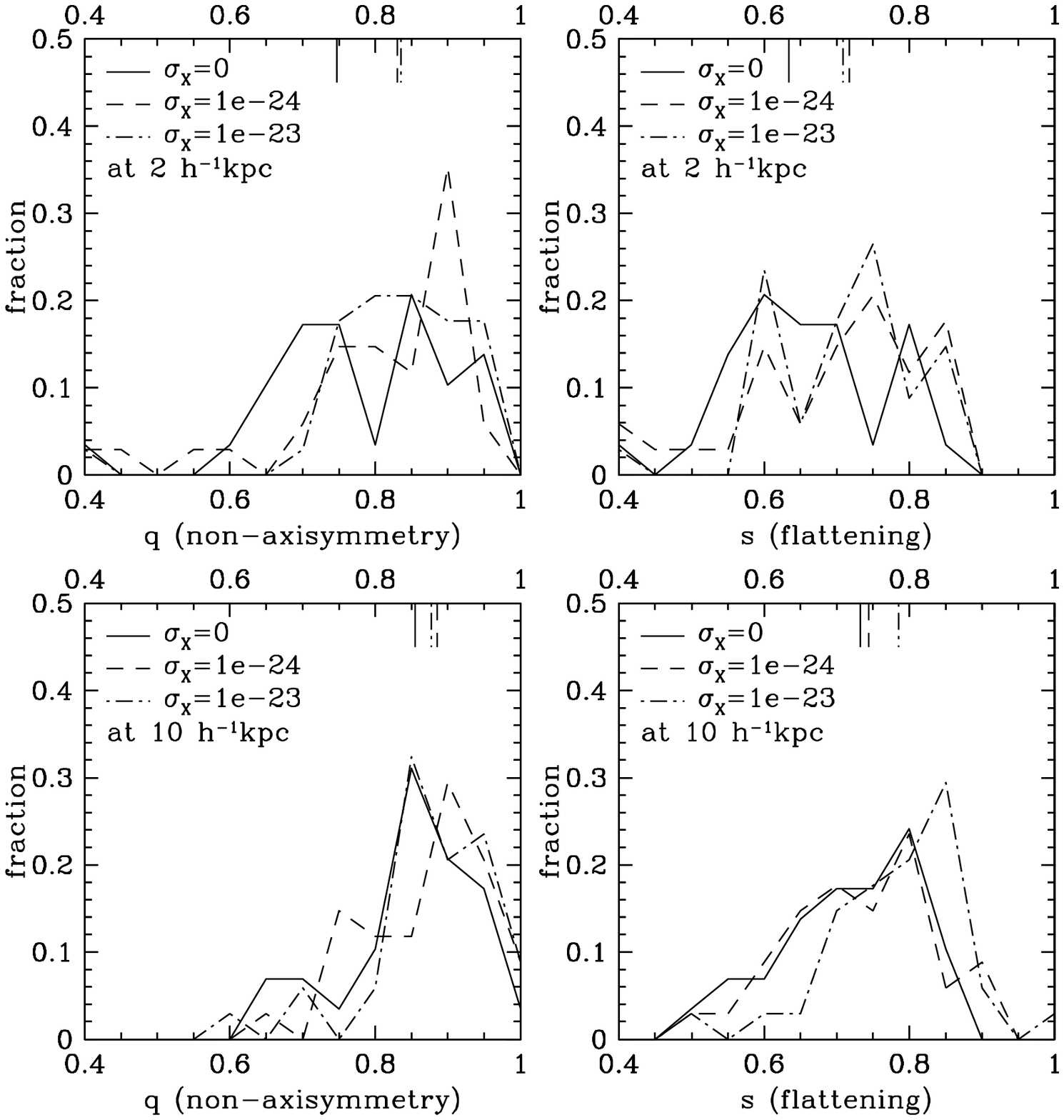 angle=0 voffset=-100 hoffset=-40
vscale=48 hscale=48}{3.0in}{3.8in} 
{\\\small Figure~9: Histogram of axis ratios $q$ (left panels) and $s$ (right
panels), at $2\hkpc$ (upper panels) and $10\hkpc$ (lower panels).
Only halos with more than 1000 particles are included.  Tick marks at
the upper axis show median values.
\vskip0.1in
}

A comparison with observations of halo shapes is as yet inconclusive.
In the inner portions of dark halos the shape of the potential is
likely to be dominated by baryons, so a comparison to these simulations
is not straightforward.  Further out, perhaps the most direct
observations of axisymmetry are those for galaxies with HI rings such
as IC~2006, which suggests a very axisymmetric halo, $q\approx 0.93\pm
0.08$ at $\approx 13$~kpc (\cite{fra94}).  Other observations (see
\cite{sac99}) are more dependent on observational and theoretical
uncertainties such as viewing angle and potential modeling, but
persistently suggest $q\ga 0.8$ at $\sim 15-20$~kpc.  Both CDM and SIDM
halos are consistent with these observations.  Lensing maps of galaxies
and clusters offer the best hope for mapping the mass potential in the
inner halos, which should place strong constraints on SIDM.

Conversely, observations of $s$ from polar ring galaxies (e.g.
\cite{sac94}) and X-ray isophotes (e.g. \cite{buo98}) suggest a
substantial amount of flattening, $s\approx 0.5\pm 0.2$ at $r\sim
15$~kpc in the density distribution.  Such a flattening, if confirmed,
may prove troublesome for SIDM.  The baryonic component would provide a
flattened contribution, but is not expected to be significant at those
radii.  It is not immediately evident how these discrepancies may be
resolved, but we note that the problem is almost as severe for CDM as
SIDM in our simulations.  It is worth mentioning that our small simulation
volume results in significantly reduced tidal distortion of large
halos, so our simulations may not accurately represent the
ellipticities of the outer portions of halos.

\section{The Monte Carlo Resolution Limit}\label{sec: res}

Our spatial resolution and mass resolution are well-understood.  However,
another resolution issue arises due to the Monte Carlo modeling of
self-interactions.  A Monte Carlo method must be sufficiently well
sampled, resulting in a separate criterion for the number of particles
in a halo to be well-represented by our simulation technique.  In this
section, we determine this criterion using our suite of lower resolution
simulations with $64^3$ particles described in \S\ref{sec: sim}.

Since we are most concerned with the inner parts of halos, we focus on
the inner slope as a function of mass as the best measure for examining
this Monte Carlo resolution limit.  Figure~10 shows a plot similar to
right panel of Figure~4, except for the $64^3$ simulation of the
$\sdm=10^{-23}\xunits$ SIDM model.  The top axis shows the number of
particles in these halos.  Here we compute $\alpha$ at $2\hkpc$ since
that is the spatial resolution of our $64^3$ runs.  The dashed line is
the running median $\alpha$ from the $128^3$ simulation, computed at
$2\hkpc$.

Figure~10 shows that for halos with $\ga 1000$ particles in the $64^3$
run, the median value of $\alpha$ is within $1\sigma$ of that of the
$128^3$ run, though consistently lower.  By 300 particles, the value of
$\alpha$ is significantly lower in the $64^3$ run.  The reason it is
lower is because with few particles, the Monte Carlo procedure results
in too few interactions to make the profile depart significantly from
the collisionless CDM case.  Thus to lower masses, SIDM looks
increasingly like CDM when modeled using this technique.

\PSbox{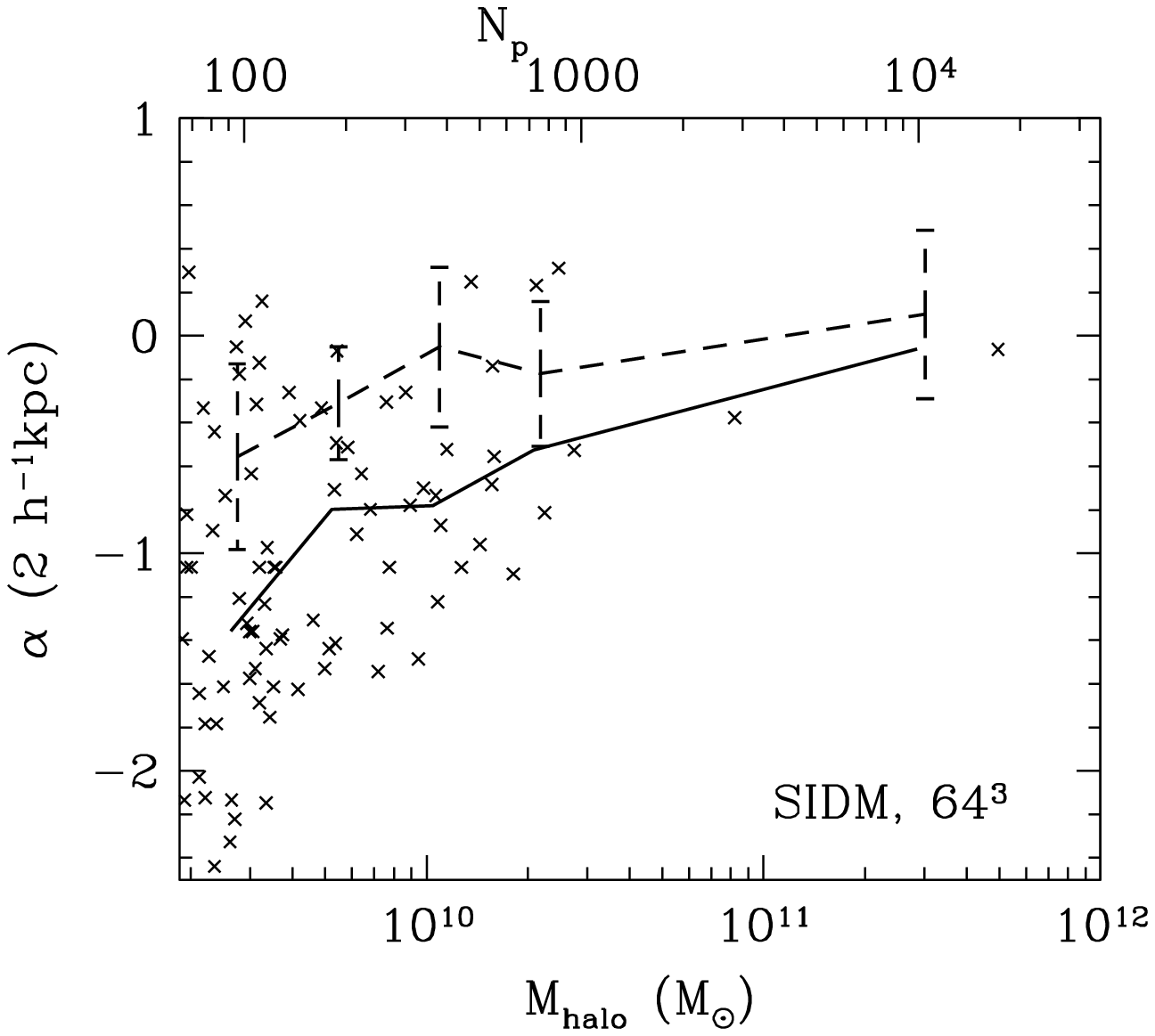 angle=0 voffset=-200 hoffset=-40
vscale=48 hscale=48}{3.0in}{2.3in} 
{\\\small Figure~10: $\alpha$ vs. $M_{\rm halo}$ for SIDM with
$\sdm=10^{-23}\xunits$, for our $64^3$ run.  $\alpha$ here is measured
at $2\hkpc$, the resolution of our $64^3$ runs.  Solid line shows a
running median of the $\alpha$ distribution.  Dashed line with error
bars shows a similar curve from $128^3$ run.  Deviations between two
at a level $>1\sigma$ occur for halos having somewhere between 300 and 1000 particles.
\vskip0.1in
}

Note that this limit is specific to our simulation parameters, redshift,
and $\sdm$, and is not a general statement about the Monte Carlo
$N$-body technique.  The limit becomes higher as $\sdm$ is lowered, since
collisions become less frequent, but even for $\sdm=10^{-24}\xunits$,
$1000+$ particle halos also appear convergent.  We therefore take our
Monte Carlo resolution limit to be $\ga 1000$ particles.

We list the values of $\alpha_{\rm med}$ and $c_{M,\rm med}$ for the
$64^3$ runs in Table~1 for comparison with the $128^3$ results, where
the median values here are computed for all halos with more that 500
particles in these smaller runs (roughly 10 in each simulation).  Note
that $\alpha_{\rm med}$ is computed at $2\hkpc$ instead of $1\hkpc$,
partially explaining the steeper slopes even at the highest masses.  In
general, the trends indicated by the $128^3$ runs are reproduced at
this lower resolution, suggesting that discrete particle effects do not
significantly affect our conclusions.  We have also examined the
$128^3$ statistics presented previously using a limit of 300 particles
instead of 1000, and our overall conclusions remain the same.

\section{Subhalo Population}\label{sec: subhalo}

Self-interacting dark matter is predicted to significantly lower the
population of subhalos orbiting around large halos, thereby bringing
simulation predictions into better agreement with observations of the
Local Group dwarf population.  There are two reasons why SIDM has this
effect:  (1) The lowered central concentration and larger core radius
makes small halos more susceptible to tidal disruption, and (2) Dark
matter is ram-pressure stripped out of small galaxies as they move
through the large central halo.  In our $128^3$ simulations, we have one halo
that is roughly Milky Way sized, having $M\approx 6.7\times 10^{11}
M_\odot$.  In this section we examine the subhalo population around
this large halo.

\PSbox{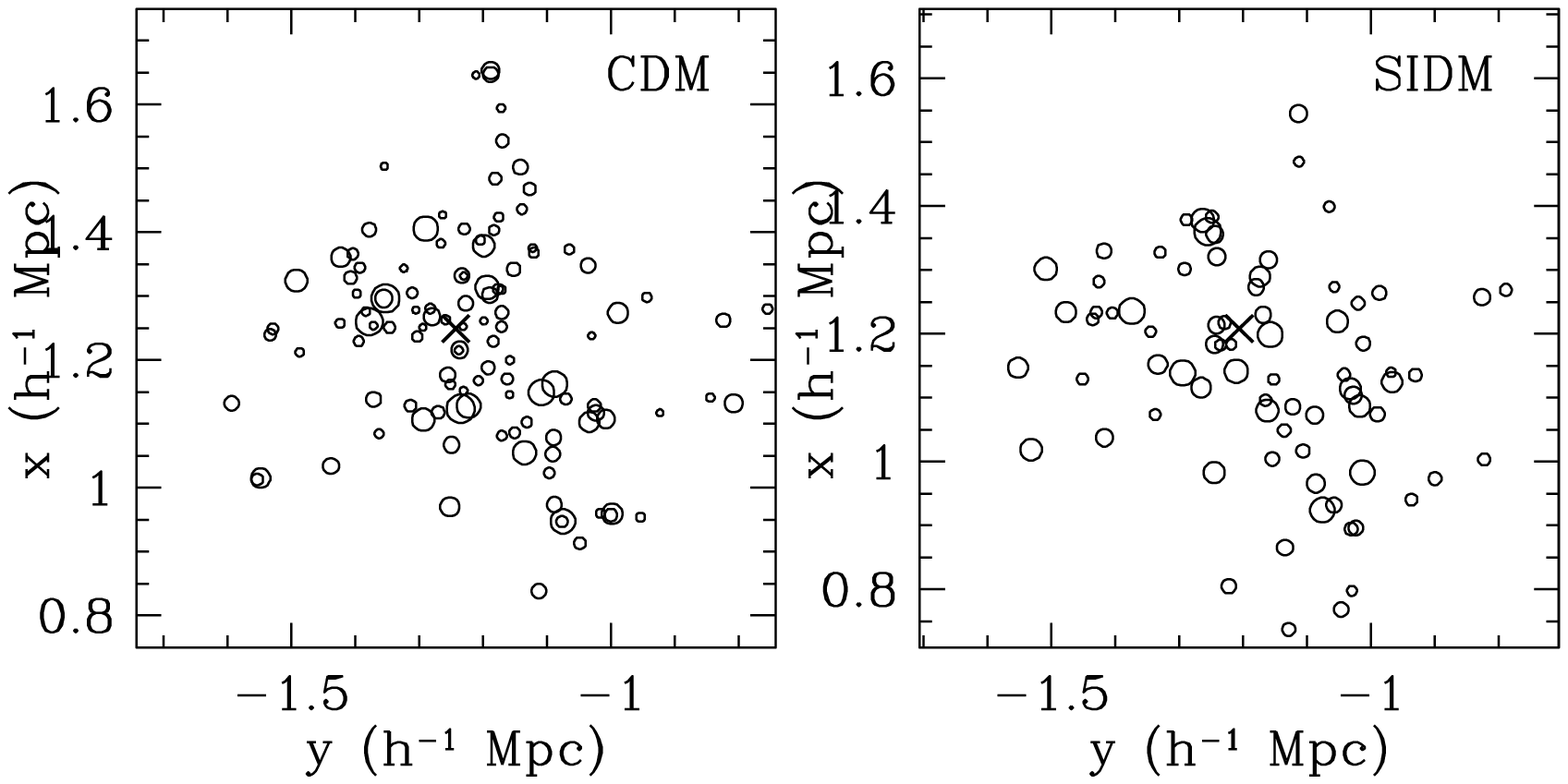 angle=0 voffset=-220 hoffset=-40
vscale=48 hscale=48}{3.0in}{2.0in} 
{\\\small Figure~11: Subhalo positions within $500\hkpc$ the largest
object in our volume, in CDM (left panel) and SIDM with
$\sdm=10^{-23}\xunits$ (right panel).  Central cross is the position of
the large halo.
\vskip0.1in
}

Figure~11 shows a projected plot of halos within $500\hkpc$ of the
largest halo in our volume, indicated by the central cross.  Circle
sizes are scaled as $\log(M_{\rm halo})$, with the smallest circles
representing halos with $M\approx 5\times 10^8 M_\odot$.  Left panel
shows CDM, right panel shows SIDM with $\sdm=10^{-23}\xunits$.  The
positions of subhalos are different due to the accumulated differences
of chaotic orbits within a highly nonlinear potential well.  Thus a
halo-by-halo comparison for these small halos is not 
possible.  A careful examination reveals that SIDM has fewer subhalos
than the CDM distribution, especially the smallest ones.

\PSbox{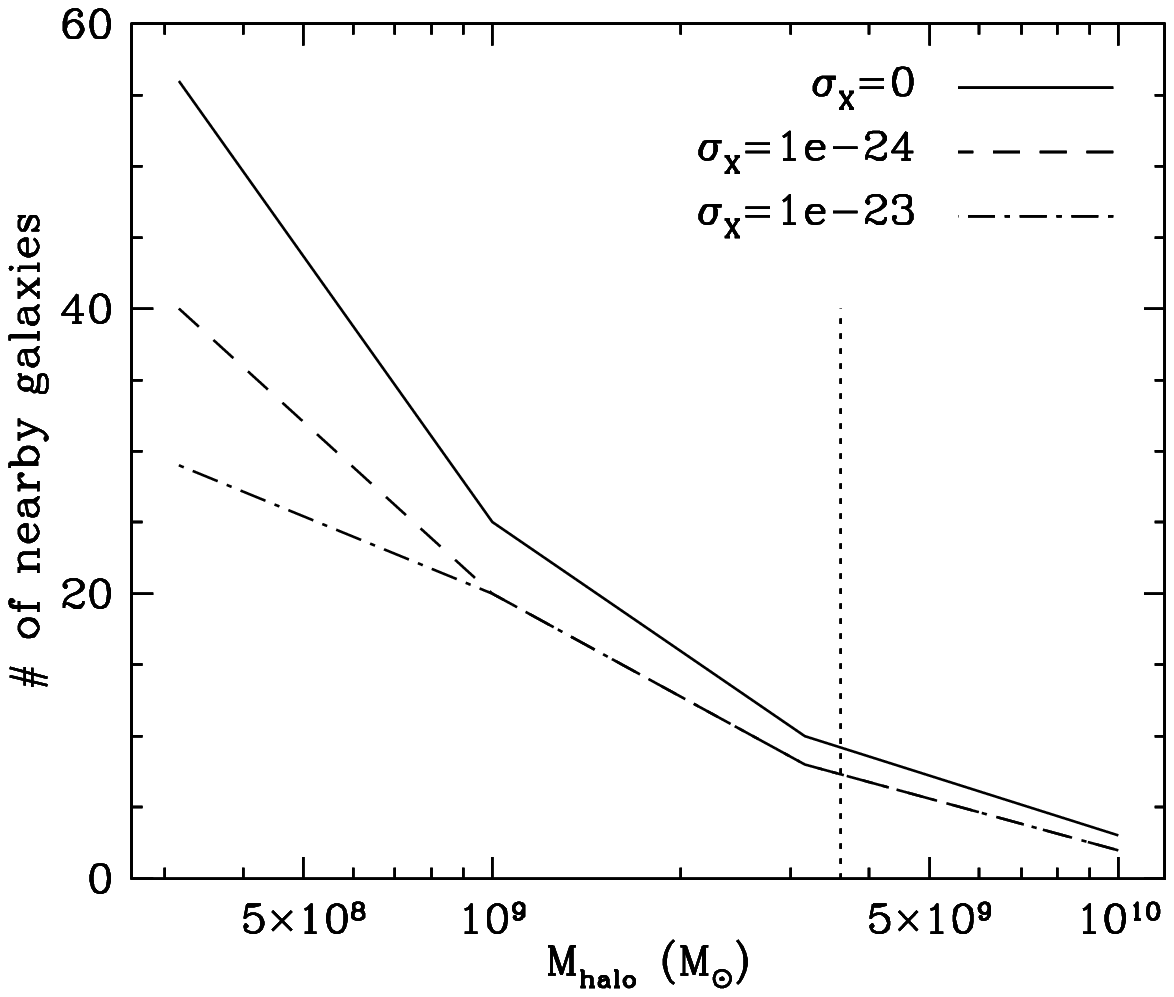 angle=0 voffset=-200 hoffset=-30
vscale=48 hscale=48}{3.0in}{2.3in} 
{\\\small Figure~12: Number of halos within $500\hkpc$ of the largest
halo, histogrammed by mass, for collisionless (solid),
$10^{-24}\xunits$ (dashed), and $10^{-23}\xunits$ (dot-dashed).
Vertical dotted line indicates our 1000-particle Monte Carlo resolution
limit.
\vskip0.1in
}

Figure~12 quantifies this effect, showing the mass function of halos
within $500\hkpc$ of our largest halo, for our three $128^3$
simulations.  There is a clear trend that SIDM suppresses the subhalo
population at the smallest masses.  For $10^{8.5}<M<10^9 M_\odot$, CDM
has 56 neighboring halos, $\sdm=10^{-24}\xunits$ has 40, and
$\sdm=10^{-23}\xunits$ has 29.

While SIDM does reduce the population of smallest halos, the reduction
is not nearly by the order of magnitude required to obtain agreement
with Local Group dwarf galaxy counts (\cite{moo99a}).  However, the
effect of subhalo suppression in these SIDM simulations should be
regarded as a lower limit to the true strength of the effect.  The
reason is that almost all these subhalos are well below our Monte Carlo
resolution limit (dotted line in Figure~12), therefore their
concentrations are approaching those in CDM models.  Thus tidal
disruption of these halos is not much stronger in SIDM as in CDM, and
is increasingly similar to lower masses.  Furthermore, ram-pressure
stripping is reduced in effectiveness for the same reason that the
Monte Carlo technique is less effective in these small halos.  Hence
the numbers of small SIDM halos are not as significantly suppressed
relative to CDM as they should be.

We conclude that the simulations considered here suggest a weak trend
in reducing the number of subhalos with increasing $\sdm$, but due to
resolution effects we can make no robust quantitative estimates.  What
is required is to simulate a large halo with incredibly high
resolution, having subhalos containing thousands of particles to
properly model the effects of self-interactions.  Such a simulation is
unfortunately beyond the scope of our current computational resources.
Alternatively, more sophisticated algorithms are necessary to model
self-interactions in small halos moving through large ones, which is an
avenue we are currently pursuing.

\section{Comparison with Previous Work}\label{sec: comparison}

A number of authors have investigated SIDM using $N$-body simulations.
The literature divides into two subsets:  Those that model interactions
in the fluid approximation, effectively employing a large cross section,
and those that model self-interactions in the optically thin regime as
suggested by \cite{spe00}.  In both cases, there is disagreement over
whether SIDM makes halos less or more concentrated than CDM.

\cite{moo00} and \cite{yos00} simulate a galaxy cluster within a
cosmological context, using a treecode with SPH to model interactions
in the fluid approximation.  Both studies resulted in halos that had
isothermal profiles and were more centrally concentrated than CDM
halos.  This may be because the large effective cross section increases
heat transfer efficiency, though as mentioned in the Introduction, for
a sufficiently large cross section one expects heat transfer to be
diminished.  \cite{yos00} suggested that intermediate cross sections
would likely yield results that were intermediate between steep CDM
profiles and steeper isothermal profiles, and thereby argued against
SIDM.  However, this is contradicted by \cite{yos00b}, as well as the
results presented here, confirming that the intermediate case results
in halos that have long-lived shallow profiles.

In contrast, Bryan (private communication) uses a adaptive mesh
hydrodynamics code to model self-interactions in a cosmological volume,
and finds that even in the fluid limit SIDM produces sizeable, long-lived
cores.  It is not clear why different hydrodynamic codes give different
results when they should be operating in the same regime.  Perhaps the
effective cross section is larger in the adaptive mesh code due to
algorithmic differences, reducing heat transfer.  Another possibility
is that a numerical effect in SPH in which cold clumps moving through
hot halos have their drag significantly overestimated (\cite{tit99})
makes objects rapidly sink into a dense, isothermal core.  It is beyond
the scope of this paper to resolve these issues.  We simply note that
the highly optically thick limit is not the relevant scenario to test
the cross section range proposed by \cite{spe00}.

\cite{bur00} (2000) and \cite{koc00} (2000) simulated isolated halos
with SIDM having a cross section closer to the range of \cite{spe00}.
They begin with a fully formed cuspy galaxy halo and study the
evolution after interactions are turned on.  They both find that halos
develop a shallow core for some length of time, and then undergo core
collapse.  \cite{bur00} and \cite{koc00} disagree on the timescales of
core collapse; \cite{bur00} finds $t_c\sim 16t_{\rm dyn}$ in agreement
with estimates from two-body relaxation, while \cite{koc00} finds a
much shorter collapse timescale of $t_c\sim 2t_{\rm dyn}$ for the same
dark matter cross section.  \cite{koc00} explain this difference by
arguing that \cite{bur00}'s method underestimates collisions of slow
moving particles.  We note that our method does not suffer from this
concern, as it is more like \cite{koc00}.

We suggest some possible reasons why the results of \cite{koc00} and
(to a lesser extent) \cite{bur00} are at odds with ours.  The first is
that they begin with a cuspy Hernquist profile.  This halo evolves
rapidly initially (as seen in Figure~3 of \cite{koc00}), resulting in
an artificially large amount of heat transfer.  Second, they simulate
an isolated halo, ignoring the accretion of dynamically hot material
during the formation process that would keep the outer halo hot and
delay core collape.  We note that their dimensionless cross section
$\hat\sigma_{DM}=M_{\rm halo}\sdm/r_s^2$ converts to ours by a factor
of $\sim 2\times 10^{-23}\xunits$ for $M_{\rm halo}\approx 2\times
10^{10}M_\odot$ and $r_s\approx 5\hkpc$ (cf. Figure~1).  \cite{koc00}'s
simulation with $\hat\sigma_{DM}=1$ already produces halos that
maintain cores over many dynamical times (cf. top panel of their
Figure~2).  We suggest that a somewhat smaller $\hat\sigma_{DM}$ might
be consistent with observations as well as their limits on core
collapse timescales, even without considering the effects of accretion
and merging.

\cite{koc00} also point out, as we have, that the Monte Carlo $N$-body
technique requires a large number of particles for accurate modeling,
and show that $10^5$ particles is sufficient.  We note that while most
of our halos do not have that many particles, our largest (Milky Way
sized) halo has roughly $2\times 10^5$ particles, and its properties
are consistent with those of smaller halos.

\cite{yos00b} have now performed a cosmological cluster simulation with
a cross sections $\sdm\approx 2\times 10^{-25}-2\times 10^{-23}\xunits$
using a Monte Carlo $N$-body method.  They find shallower central
slopes and less concentrated cores with SIDM.  Our combined results
span the range from dwarf galaxies to clusters, and are in broad
agreement with each other if we focus purely on the numerical results.
For instance, they find that $\sdm\approx 2\times 10^{-23}\xunits$
produces a cluster having a core size of $160\hkpc$.  If we extrapolate
our largest halo's core size (in our $\sdm=10^{-23}\xunits$ run) to
cluster scales using the expected $r_{\rm core}\propto v_{\rm circ}$,
we predict a core size of $\sim 150\hkpc$ for their cluster.  For
$\sdm\approx 2\times 10^{-24}\xunits$, they get $100\hkpc$ core, while
we would predict $\sim 80\hkpc$ for $\sdm=10^{-24}\xunits$, again in
good agreement.  They then proceed to scale down from clusters to dwarf
galaxies using a simplistic argument based on number of collisions, but
their scaled result is contradicted by our direct simulations.  We
estimate that this may be because they use a CDM value for the scale
radius and $\cnfw$ of dwarfs, and compare them to SIDM values for their
cluster.  We find that $\sdm=10^{-24}\xunits$ produces \cite{nfw96}
scale radii that are double that of CDM (cf.  Figure~6 and discussion);
such a factor would go a long way towards alleviating the discrepancy.
Taking this into account, we find the simulations of \cite{yos00b} to
be broadly consistent with ours.

\PSbox{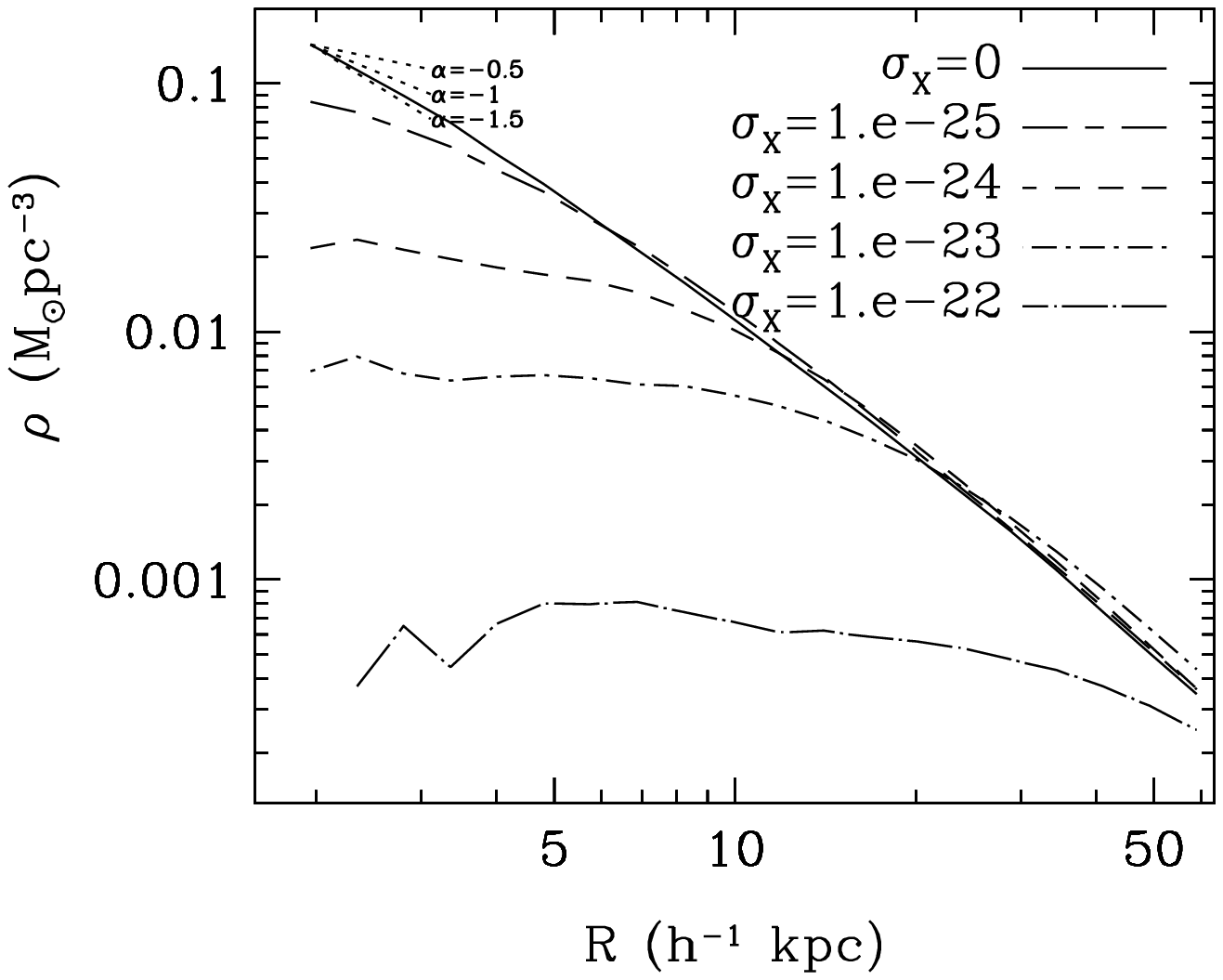 angle=0 voffset=-200 hoffset=-40
vscale=48 hscale=48}{3.0in}{2.1in} 
{\\\small Figure~13: Halo profile of the largest halo in our $64^3$
simulations, for a range of $\sdm$ values.  Halos are progressively
less concentrated and have larger cores with increasing $\sdm$.
\vskip0.1in
}

In order to explore the high-$\sdm$ limit, we ran $64^3$ simulations of
SIDM with $\sdm=10^{-25}-10^{-22}\xunits$.  The most illustrative
result is to compare the density profile of the largest halo in all our
$64^3$ simulations, as shown in Figure~13.  As seen in Figure~1, there
is a smooth trend of increasing core radius with $\sdm$.  SIDM with
$\sdm=10^{-25}\xunits$ is quite similar to CDM, though it may also have
a core below our $2\hkpc$ resolution limit.  Increasing $\sdm$ to
$10^{-22}\xunits$, we continue to see no evidence for the development
of an isothermal core due to accelerated heat transfer.  The reason is
because the collisions are so frequent in the outer portion of the halo
that a dense core cannot develop.  Instead, collisions randomize the
dark matter velocities and prevent a smooth radial inflow required to
generate a dense core.  As dynamically hot material accretes onto the
halo, heat keeps flowing inward and  a large core is maintained.  Our
results are in better agreement with Bryan as opposed to \cite{moo00}
and \cite{yos00}.  This also illustrates why simulating SIDM beginning
with an isolated cuspy Hernquist profile may not be appropriate for
large $\sdm$; one should at least begin with a halo profile that is
self-consistently stable for a few dynamic times.

\section{Summary}\label{sec: disc}

We present a set of cosmological self-interacting dark matter
simulations having cross-sections in the range favored by \cite{spe00}
(2000).  Our simulations include the growth of halos from linear
fluctuations in a random volume of the universe, with sufficient volume
and resolution to obtain a statistical sample of galactic halos
resolved to $1\hkpc$.  We compare the resulting halos on a case-by-case
basis to those in a collisionless CDM simulation having the same
initial conditions.

Overall, SIDM is remarkably successful at reproducing observations
of the inner portions of dark matter halos where CDM appears to fail.
In particular, we find:

\begin{enumerate}

\item The inner slopes of SIDM with $\sdm =10^{-23}\xunits$ typical
halos have $\alpha\approx -0.4$ at $r\sim 1\hkpc$, with some scatter in
$\alpha$.  Our CDM halos have $\alpha\approx -1.5$, in agreement with
previous studies (e.g. \cite{moo99} 1999).  SIDM with $\sdm
=10^{-24}\xunits$ is intermediate between these cases, with median
$\alpha\approx -0.9$.  SIDM is in better agreement with a preliminary
analysis of H$\alpha$ rotation curves of low surface brightness
galaxies (\cite{dal00}).

\item SIDM with $\sdm =10^{-23}\xunits$ produces central densities
$\rho_c\sim 0.01\Mpc3$ at $1\hkpc$, and shows no trend with halo mass.
SIDM with $\sdm =10^{-24}\xunits$ has somewhat higher $\rho_c$ values,
but remains fairly independent of mass.  Conversely, $\rho_c$ in CDM
halos is much larger than observed, typically $\ga 0.1\Mpc3$ at
$1\hkpc$, and shows a strong trend with halo mass.  With their steep
profiles, CDM halos are in significantly worse agreement at smaller radii.
SIDM is thus is in better agreement with observations, as has also been
argued by \cite{fir00a}.

\item Simulations with SIDM having $\sdm =10^{-24}\xunits$ are
intermediate between CDM and SIDM with $\sdm =10^{-23}\xunits$,
indicating a smooth increase in the effect of SIDM with cross section,
a result that extends (using lower-resolution simulations) from $\sdm
=10^{-25}\rightarrow 10^{-22}\xunits$.  In particular, the
generation of singular isothermal halos is not seen in any of the
massive halos simulated, even for $\sdm =10^{-22}\xunits$.   This
suggests that the dynamical process of halo growth in a cosmological
setting helps keep outer regions of halos hot and prevents core
collapse in a Hubble time.

\item We introduce a new mass concentration parameter $c_M$ based on a
more directly observable quantity, the enclosed mass within tens of
kpc.  This halo concentration is significantly lower in SIDM models as
compared to CDM, providing an observationally accessible discriminant
that is not dependent on fitting a particular profile form.  A rough
estimate of $c_M$ for the Milky Way, with large corrections for
baryonic effects, favors SIDM over CDM.

\item The central phase space density is lower in SIDM vs. CDM mostly
due to the reduction in $\rho_c$.  The velocity dispersions in the
inner regions are quite similar.  Both SIDM and CDM are consistent with
observations shown in \cite{dal00a}, though SIDM is mildly favored.

\item SIDM produces halos that are more spherical, especially in their
inner regions, as compared to CDM.  In principle, this is one of the
strongest tests of the SIDM paradigm, as near the center {\it any}
value of $\sdm$ that has a non-negligible effect on the dark matter
distribution will increase the core sphericity, while CDM cores are
almost always significantly triaxial.  However, baryons are likely to
dominate the shapes of the inner parts of halos, complicating a direct
comparison, and in the outer parts the differences between SIDM and CDM
are less pronounced.

\item The number of subhalos around our largest (Milky Way-sized) halo is
somewhat reduced with increasing $\sdm$, but due to discreteness effects
in our Monte Carlo $N$-body technique, we cannot put robust quantitative
estimates on the strength of this effect.

\end{enumerate}

Based on these simulations, our currently favored value for $\sdm$ is
somewhere between $10^{-23}\xunits$ and $10^{-24}\xunits$.  Such a
cross section simultaneously reproduces both the observed central
density and inner slope, as well as being consistent with various
observations considered here, which is non-trivial.  In contrast, for
instance, warm dark matter has difficulty simultaneously reproducing
the observed central densities, inner slopes and subhalo population
(\cite{col00}).

As stated before, inner halo shapes may provide a strong discriminant
between CDM and SIDM.  On galactic scales, they are difficult to
observe and confused by baryonic contributions.  Conversely, clusters
provide a cleaner test because they have large cores that are not
baryon-dominated, and their mass distributions are directly observable
via lensing.  \cite{mir00} uses the asphericity of cluster MS~2137-23
to (analytically) argue that $\sdm<10^{-25.5}\xunits$, effectively
ruling out SIDM as a solution to halo concentration problems.  On the
other hand, CL~0024+1654 is very spherical, much more so than CDM
models generally predict (\cite{tys98}).  Our simulations cannot
directly address the shapes of clusters, as we have no cluster-sized
objects in our volume.  However, SIDM shows some range of halo shapes
due to asymmetric infall and unrelaxed mass distributions, so it is
unclear whether a single object can definitively rule out SIDM.
Support for this statement is provided by \cite{yos00b}, whose cluster
has enough triaxiality to be consistent with MS~2137-23 even for
$\sdm\approx 2\times 10^{-24}\xunits$ (their model S1Wb), contradicting
Miralda-Escud\'e's scaling argument.  We note that halo shapes are
unaffected by annihilating (\cite{kap00}) or decaying (\cite{cen00})
dark matter, thus they also provide a discriminant between these
variants and SIDM.

The cluster core sizes in the simulations of \cite{yos00b} are larger
than observed ($\sim 30-70\hkpc$, \cite{mir95}; \cite{tys98}),
certainly for $\sdm\approx 2\times 10^{-23}\xunits$ ($160\hkpc$), and
probably even for $\sdm\approx 2\times 10^{-24}\xunits$ ($100\hkpc$).
So it may be that SIDM has difficulty matching observations at both
dwarf galaxy and cluster scales.  However, adiabatic contraction of
baryons during the formation of the cD galaxy will reduce the cluster
core radius from $N$-body predictions, so the discrepancy may not be
that large.  In any case, SIDM with $\sdm\approx
10^{-24}-10^{-23}\xunits$ comes remarkably close to matching dwarf
galaxies, $L^\ast$ galaxies and clusters given the $\sim 10^5$ range in
mass scales, so we reserve judgement pending a more careful comparison
with observations.  \cite{yos00b} mention that SIDM core sizes would be
in better agreement with observations if $\sdm\propto v^{-1}$, which
would result in the effects of self-interactions being diminished in
hot cluster environments as compared to galaxies.  Such a scenario
occurs naturally if the dark matter--dark matter scattering has
low-lying resonance or bound state contributions, as is the case for
ordinary nucleons.

Another theoretical avenue explored in relation to SIDM has been
modeling the Tully-Fisher relation.  Hydrodynamic simulations show that
the simulated Tully-Fisher zero point may be brought into agreement
with observations only if halos are less centrally concentrated than
predicted by CDM (\cite{nav00}).  \cite{mo00} determine that a cross
section of $\sdm\sim 10^{-23}\xunits$ would produce a correct
Tully-Fisher relation for $v_{\rm circ}\approx 100$~km/s halos.  Thus
it is conceivable that the cross section preferred from halo structure
constraints may also alleviate the Tully-Fisher discrepancies.

\cite{ost00} suggested that dark matter interactions in the centers of
halos naturally produce central black holes with a mass scaling $M_{\rm
BH}\propto \sigma^{5}$, in agreement with observations (\cite{mag98};
\cite{fer00}), and he estimates $\sdm\la 10^{-24}\xunits$ in order to
avoid central black holes that are too large.  However, this estimate
is based on an $\rho\propto r^{-2}$ profile all the way in to the black
hole.  If such a dense center never arises because collisions inhibit
its formation, then the limits on $\sdm$ are weakened considerably.

Observations of dark matter halos promise to improve significantly in
the coming years, particularly constraints on halo core shapes from
lensing and on inner profiles and concentrations of halos from
H$\alpha$ rotation curves.  If the inner parts of dark matter halos are
found to be generically triaxial, this would be the high place of
sacrifice for self-interacting dark matter; conversely, spherical halos
would provide strong support for this scenario.  The main modeling work
yet to be is done is an improved examination of the subhalo populations
in SIDM, as well as simulations of a larger range of mass scales.  The
$N$-body Monte Carlo approach has difficulty achieving a large dynamic
range due to the stringent Monte Carlo resolution limit (i.e.
discreteness effects in the probabalistic description of collisions),
therefore a different approach may be necessary.

The SIDM simulations presented here are a first attempt at examining
the effect of self-interacting dark matter within the context of a
realistic halo formation scenario.  The results are quite encouraging
that this simple variant of the cold dark matter paradigm will
alleviate a wide range of difficulties faced by CDM on galactic
scales.  We look forward to further investigations and comparisons with
observations.

\bigskip
\acknowledgments

We thank Julianne Dalcanton, Lars Hernquist, Jerry Ostriker, Penny
Sackett, Scott Tremaine, Martin White, and Naoki Yoshida for helpful
discussions.  We thank Greg Bryan for sharing his unpublished results.

RD and DNS are supported by NASA ATP grant NAG5-7066.
DNS and BDW are supported by the NASA MAP/MIDEX program.
PJS is supported by United States Department of Energy grant DE-FG02-91ER40671.

\end{document}